\documentclass[onecolumn,notitlepage,floatfix,amsmath,amssymb,longbibliography]{revtex4-1}

\usepackage{graphicx}
\usepackage{dcolumn}
\usepackage{color}
  		
\usepackage{amssymb}
\usepackage[toc,page]{appendix}

\newcommand{\be}{\begin{equation}}
\newcommand{\ee}{\end{equation}}

\usepackage{graphicx}
\usepackage[percent]{overpic}
\usepackage{pict2e}

\begin{document}
\title{Statistics of single and multiple floaters in experiments of surface wave turbulence}

\author{Nicol\'{a}s F. Del Grosso, Luc\'{i}a M. Cappelletti, Nicol\'{a}s E. Sujovolsky, Pablo D. Mininni, and Pablo J. Cobelli}
\affiliation{
  Universidad de Buenos Aires, Facultad de Ciencias 
  Exactas y Naturales, Departamento de F\'\i sica, \& IFIBA, CONICET, 
  Ciudad Universitaria, Buenos Aires 1428, Argentina.}

\begin{abstract}
We present laboratory experiments of surface wave turbulence excited by paddles in the deep water regime. The free surface is seeded with buoyant particles that are advected and dispersed by the flow. Positions and velocities of the floaters are measured using particle tracking velocimetry. We study the statistics of velocity and acceleration of the particles, mean vertical displacements, single-particle horizontal dispersion, and the phenomenon of preferential concentration. Using a simple model together with the experimental data, we show that the time evolution of the particles has three characteristic processes that dominate the dynamics at different times: drag by surface waves at early times, trapping by short-lived horizontal eddies at intermediate times, and advection by a large-scale mean circulation at late times.
\end{abstract}
\maketitle

\section{Introduction}
The disordered state of waves observed in the ocean surface is often considered a paradigmatic manifestation in nature of wave turbulence \cite{zakharov12}, and has been a key motivation in the last decades for the development of weak turbulence theories. In these theories, which have found multiple applications in fluid dynamics as well as in other areas \cite{zakharov12, newell11, nazarenko11}, a self-similar energy spectrum and turbulent behavior arises as a result of weak nonlinear interactions between waves of small amplitude. Recent laboratory experiments in the weak turbulent regime have shed new light on this problem, probing the dispersion relation of the waves, as well as the nature of the nonlinear coupling, in problems that range from bending waves in thin plates \cite{cobelli09, mordant10}, to gravity-capillary waves in free surface flows \cite{cobelli11, berhanu13, aubourg15, clark15, deike2015role}. While in these laboratory experiments surface waves are often generated using servo-controlled paddles or ``wavemakers,'' more recently waves have also been excited using horizontal winds \cite{paquier15, paquier16}, generating laboratory conditions more akin to those found in oceanic and geophysical flows. For a range of parameters, these studies measured dispersion relations that remain close to the theoretical linear dispersion relations of the systems, confirmed the development of turbulence in the cases in which the flows were stirred with paddles, and in some cases found energy spectra compatible with those predicted by weak turbulence theory.

In natural scenarios in which turbulence manifests itself, transport processes (such as the mixing of passive scalars, of Lagrangian and inertial particles, and of floaters in the ocean) are often dominated by the convective action of the velocity fluctuations. These processes are generally studied in homogeneous and isotropic turbulence (see, e.g., \cite{mordant01} and references therein), where they are known lo lead to particle dispersion and diffusion \cite{rast11, rast16}, and to the so-called preferential concentration of particles \cite{salazar08}: particles with different density than the fluid have the tendency to distribute inhomogeneously in space, accumulating in clusters which have been widely observed in experimental and numerical studies \cite{goto08, coleman09, obligado14, uhlmann_clustering_2017}. However, in the presence of waves there is little knowledge of how waves and eddies couple to affect particle diffusion and dispersion (see, e.g., a discussion in the context of wave turbulence in \cite{balk2004growth}). Recently, the dispersion of Lagrangian particles \cite{aartrijk08, sujovolsky_single-particle_2017, sujovolsky_vertical_2018}, and the clustering of neutrally buoyant inertial particles and of floaters \cite{van_aartrijk_vertical_2010, sozza16, sozza2018inertial} have been studied in numerical simulations of the bulk of stably stratified flows, which display a mix of turbulent eddies and internal gravity waves. However, with the exception of some recent laboratory experiments studying particles in Faraday waves \cite{francois2014three, punzmann2014generation} and clustering of floaters in standing waves or in random superpositions of waves \cite{denissenko2006waves}, available measurements of buoyant particles in free surface flows come mostly from oceanic observations.

There are several types of measurements of the velocity and the acceleration in oceanic flows. Eulerian measurements as those obtained from, e.g., acoustic Doppler profilers, which can resolve turbulent fluctuations and, consequently, give access to vertical structure functions and other relevant turbulent characteristic quantities \cite{thomson_energy_2009}. However, measurements in the ocean are often done using buoys and floaters which, for motions with typical horizontal scales sufficiently larger than the size of the floater (with sizes between $30$ cm to $1$ m), are believed to provide information on Lagrangian properties of the flow, such as vertical and horizontal velocities (with their variances), as well as the Lagrangian velocity spectrum of internal waves which for the vertical velocity component is compatible with the empirical Garret-Munk spectrum, a flat spectrum with a peak at the buoyancy frequency \cite{garrett_space-time_1975, dasaro_lagrangian_2000}. Another effect also reported in buoyant particles dispersed by the ocean is that of preferential concentration \cite{jacobs_ocean_2015, gutierrez_clustering_2016}. The role of surface waves in all these processes is important, and recent observational and numerical results indicate that current models of mixing in the upper-ocean, where turbulence enhances the fluxes and exchanges of momentum, heat, and moisture towards the atmosphere, need to be revised to account for the influence of the surface waves \cite{dasaro_turbulence_2014}. Laboratory experiments and numerical simulations of turbulent flows with surface waves can thus play a central role in providing new information on this problem, as some required quantities are hard to measure in oceanic observations. Moreover, recent results show that phenomena such as flow dissipation and drag (which have an important impact in the mixing and transport of particles) are highly intermittent, as shown from numerical simulations of oceanic flows even at very large horizontal scales, with regions in the flow with high dissipation dominating the energy and enstrophy budgets \cite{pearson_log-normal_2018}.

In this work we present laboratory experiments of surface wave turbulence in the deep water regime, with the surface of the fluid seeded with buoyant particles or floaters. The displacement and velocity of each floater is measured using particle tracking velocimetry (PTV). We study the statistics of the floaters' velocity and acceleration, the particles' velocity power spectrum, the mean squared vertical and horizontal displacements, and the formation of clusters of particles using a Delaunay tessellation. For the mean squared vertical and horizontal displacements, we also present a simple model that captures the main features of the experimental data: a ballistic growth of the dispersion at early times dominated by the waves, a saturation at intermediate times in the horizontal dispersion associated to trapping of the floaters by short-lived mid-size eddies, and a growth of the horizontal dispersion at late times resulting from the advection of the floaters by a large-scale mean circulation that develops in the vessel. Together, the observations and the model allow us to disentangle the different physical effects contributing to the dispersion of floaters by the flow.

The structure of the paper is as follows: In Sec.~\ref{sec:exp} we describe the experimental setup and briefly characterize the surface wave turbulence observed in the experiments. Section \ref{sec:vel} presents results on velocity and acceleration of the floaters, including probability density functions (PDFs) and power spectra. In Sec.~\ref{sec:part} we study single particle displacements and compare the results with those obtained from a simple model built upon a random superposition of waves and a continuous-time random walk process. Section \ref{sec:clus} provides a brief study of preferential concentration of floaters in the experiments. Finally, Sec.~\ref{sec:conc} presents our conclusions.
 
\section{Experimental Setup \label{sec:exp}}

\begin{figure}
\centering
\includegraphics[width=16.5cm]{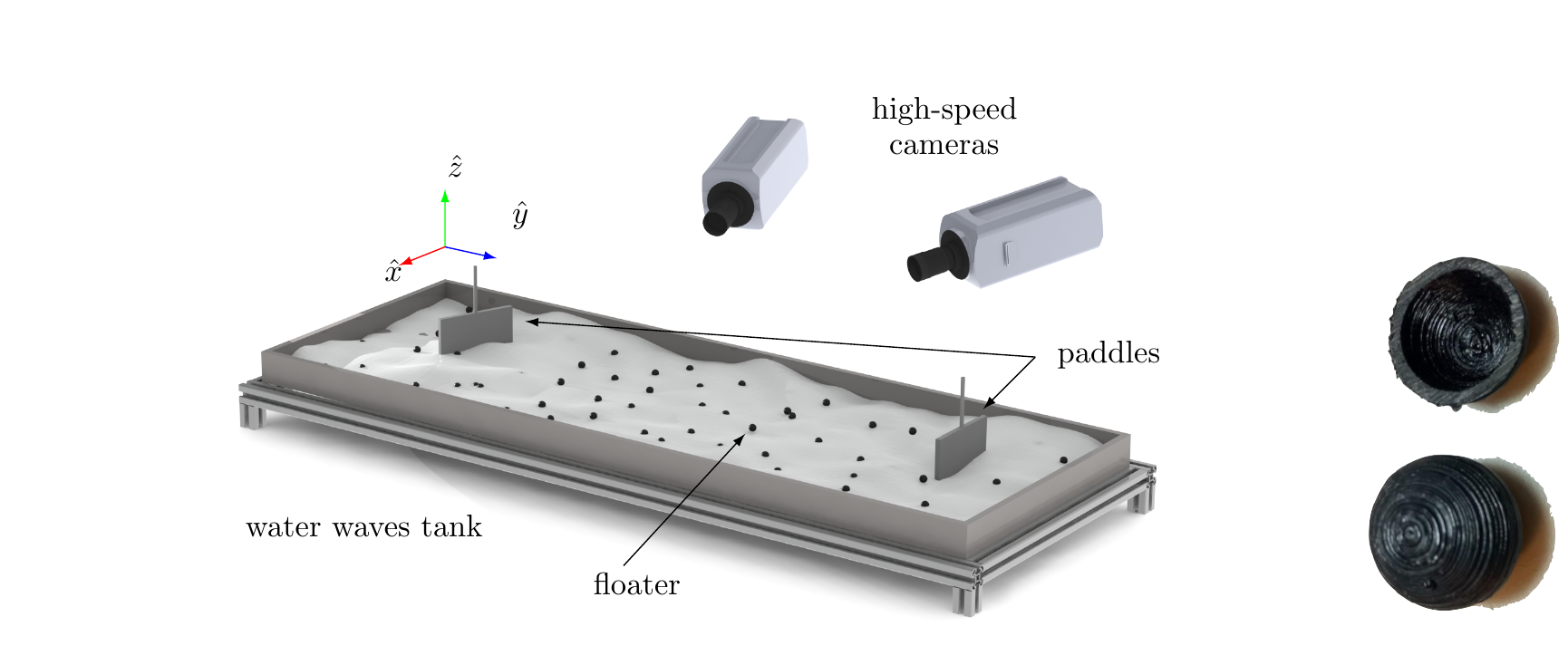}
\caption{({\it Color online}) Experimental setup. {\it Left:} the water tank with the paddles to generate surface waves, and the two cameras used for PTV. {\it Right:} a picture of a floater, with a cut showing its cross section.}
\label{f:setup}
\end{figure}

The laboratory experiments were done in an acrylic vessel (made of polymethyl methacrylate, or PMMA) of $198 \times 77 \times 36$ $\textrm{cm}^3$, with wall thickness of $1$ cm (see Fig.~\ref{f:setup}). This vessel was filled with distilled water from a double-pass reverse osmosis system to remove ions and dissolved or suspended solid particles from the water, to which we added $\textrm{TiO}_{2}$ to dye it white in order to increase contrast with the floaters. Water in the vessel was replaced regularly with water from the double-pass reverse osmosis system, to control and limit surface contamination and its possible effects on wave damping and dissipation \cite{howell2000measurements, campagne2018impact}. In all cases, the liquid column was kept at a height of $5$ cm, resulting in conditions such that waves were in a deep water regime, with dispersion relation $\omega ^{2} \approx gk \tanh(kh_0)$, where $\omega$ is the angular frequency, $g$ is the gravity acceleration, $k$ is the wavenumber, and $h_0$ is the height of the surface at rest.

A turbulent steady state of surface waves was generated and sustained by two acrylic piston-type wavemakers, driven by independently servo-controlled LinMot linear servomotors with a peak force of $47$~N and an accuracy of $0.01$ mm. The surface area of the wavemaker paddles was $15 \times 10$ cm$^2$; both plates being immersed to a depth of approximately $30$ mm. The wavemakers were driven by means of a random signal with a white frequency spectrum within the range from $0$ to $4$ Hz (as often done in experiments of gravity-capillary wave turbulence, see e.g., \cite{cobelli11}).  The signal used for the forcing thus determines a characteristic temporal scale of $\approx 0.25$~s. The same maximum amplitude $A$ for the wavemakers motion was imposed to both paddles, and its value was varied throughout the different experimental runs. In particular, experiments with values of $5$, $10$, $15$, and $20$ mm of maximum amplitude $A$ are considered in the present study, with the aim of studying regimes with different amplitude of the waves, and with different strengths of the nonlinear coupling between waves. Note the amplitude $A$ corresponds to the maximum possible amplitude of the displacement of the paddle measured from the position at rest. As a random signal in time was used, actual mean displacements were smaller (as an example, for the maximum amplitude $A$ of $10$ mm, the standard deviation of the position of the paddle from rest was of $2.7$ mm). In all cases, the choice of the forcing (and in particular, of the range of frequencies excited) was such that it reduced capillary effects in the system, and that it excited a turbulent steady state with gravity waves.

\begin{figure}
\centering
\includegraphics[width=8.9cm]{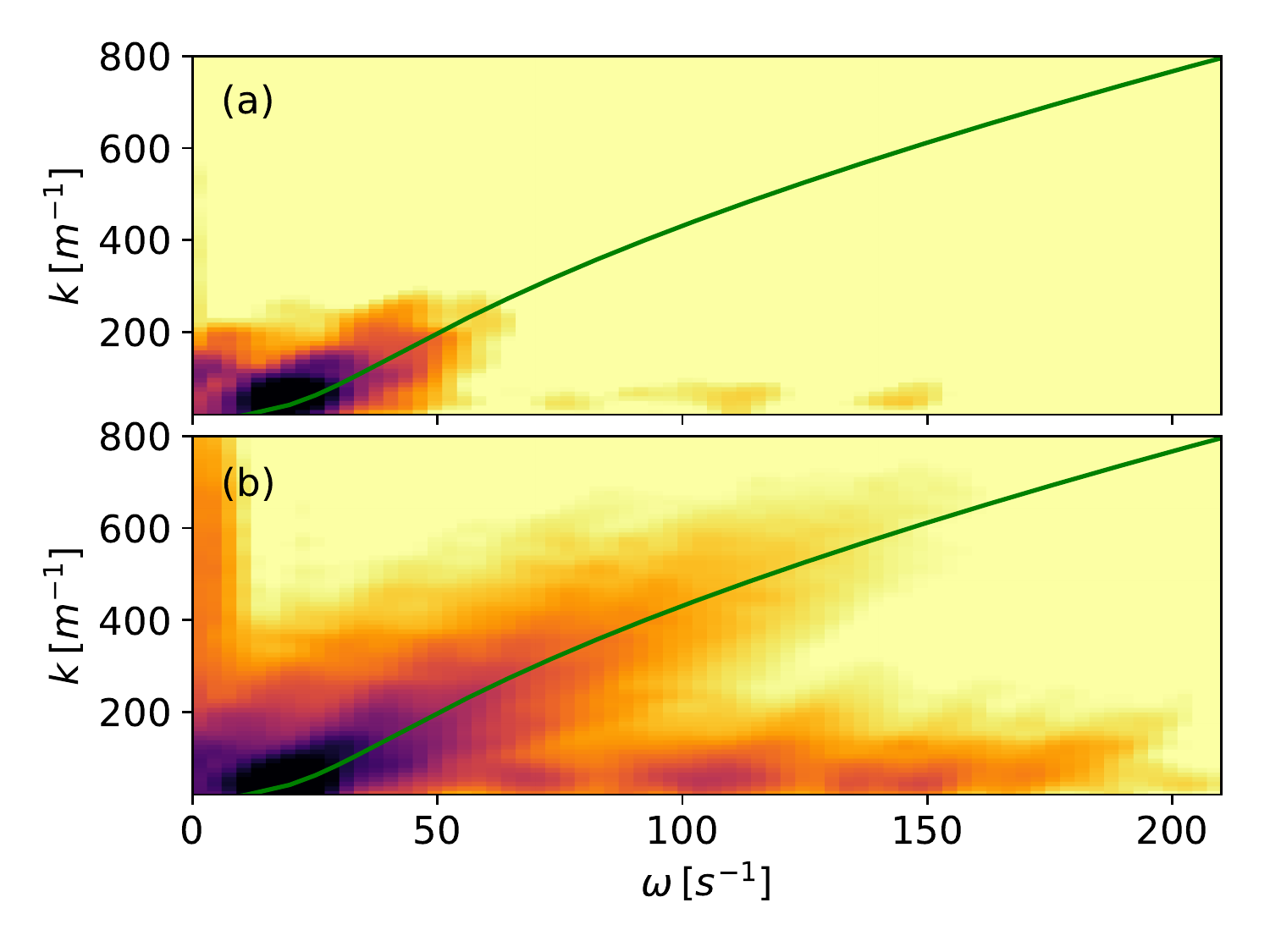}
\includegraphics[width=8.9cm]{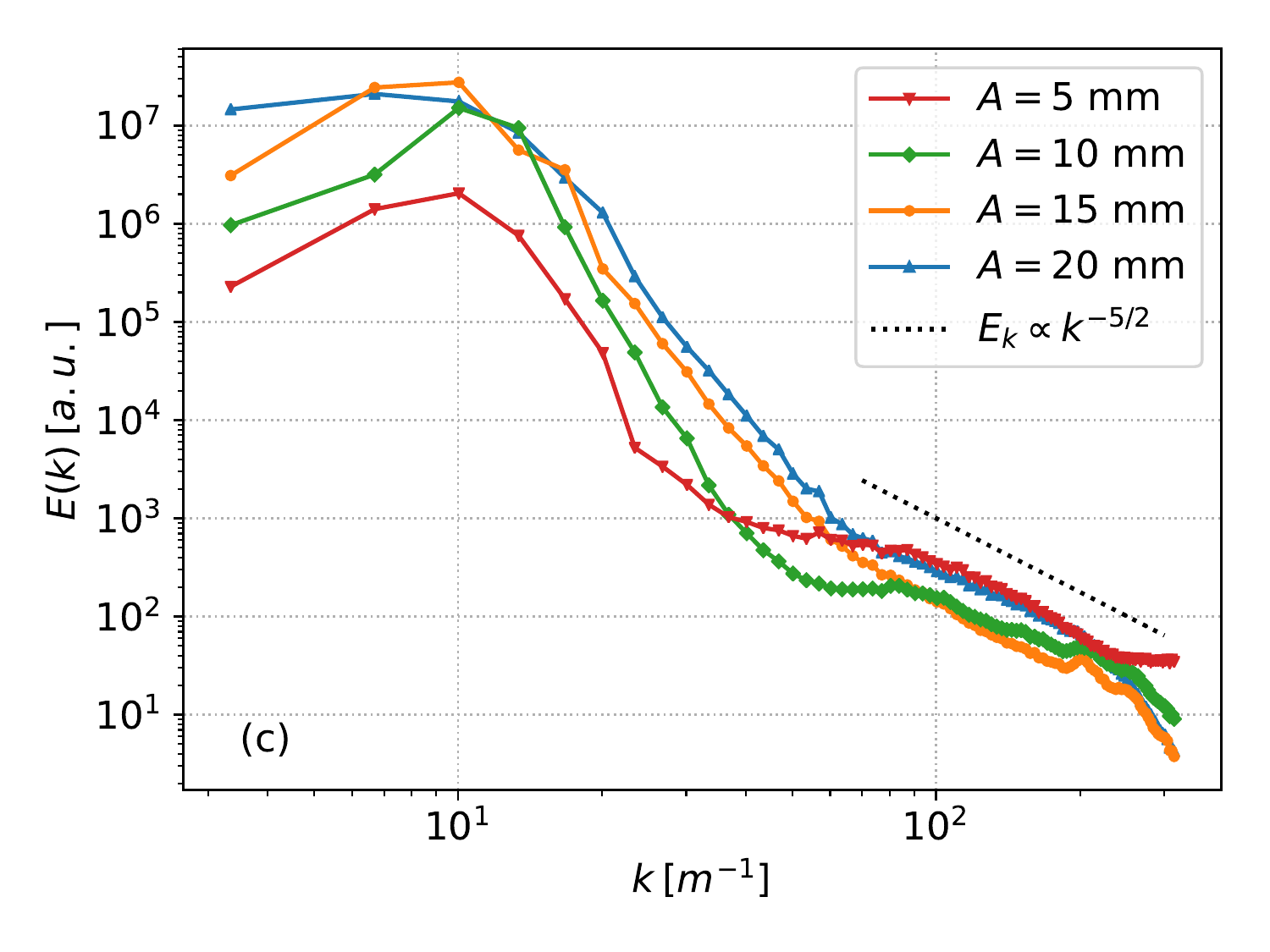}
\caption{({\it Color online}) Eulerian energy spectra of surface deformation obtained using the fringe projection profilometry technique. (a) Spatio-temporal spectrum of the potential energy $E(k,\omega)$ (or, equivalently, power spectrum of the surface deformation amplitude) for the experiments with $A=10$ cm. Darker colors correspond to larger energy densities. The green solid line indicates the theoretical dispersion relation for the system. (b) Same for the experiments with $A=20$ cm. (c) Eulerian wave number spectrum $E(k)$ of the surface deformation amplitude for all forcing functions $A$ considered in the study. An $E(k) \sim k^{-5/2}$ power law is indicated as a reference.}
\label{f:spectra}
\end{figure}

To characterize the turbulent state generated by the forcing, we performed measurements of the water surface height deformation using a fringe projection profilometry technique \cite{cobelli09} with a SA3 Photron ultra-fast camera. This technique allows us to compute the wave steepness, as well as spatio-temporal and wave number spectra of the squared surface wave height. For all values of the forcing amplitude $A$, the r.m.s.~wave steepness (which measures the strength of the non-linearity) is smaller than 3\%. Figure \ref{f:spectra} shows the spatio-temporal spectra $E(k,\omega)$ for two cases (with $A=10$ and $20$ mm), and the wave number spectrum $E(k)$ for all values of the forcing amplitude $A$. The spatio-temporal spectra show accumulation of energy in the vicinity of the theoretical dispersion relation of the waves, confirming a significant fraction of the energy in the system is in the form of wave excitations, and with a broadening associated to the strength of the nonlinearities. For $A=10$ mm, significant excitations can be seen up to $k\approx 300$ m$^{-1}$. For $A=20$ mm, a stronger excitation over the dispersion relation, reaching much larger wavenumbers (together with stronger excitations at low frequencies) can also be seen. The wave number spectra $E(k)$ in Fig.~\ref{f:spectra}(c) show a broad peak followed by a range compatible with $\sim k^{-5/2}$ scaling, from $k\approx 60$ m$^{-1}$ up to $k\approx 300$ m$^{-1}$. This power law is compatible with the solution to the Hasselmann equation for gravity waves, or equivalently, with the so-called Zakharov-Filonenko spectrum in wave number space (corresponding to a frequency spectrum $\sim \omega^{-4}$) \cite{zakharov66,nazarenko11}. A broader power spectrum (with a smaller peak at low wave numbers) is observed for the experiment with $A=5$ mm; as the forcing amplitude is increased the peak also increases and broadens, and the width of the range compatible with the power law decreases.

Interestingly, note that forcing frequencies up to 4 Hz yield (through the full dispersion relation of the system and taking into account capillary effects) forced wavenumbers up to $k_F \approx 60$ m$^{-1}$ (corresponding to wavelengths of $\approx 10$ cm). Also, considering that the crossover between the gravity and capillary waves regimes takes place at a frequency $f \approx 13.5$ Hz, we obtain an associated wavenumber for the crossover of $k_C \approx 367$ m$^{-1}$ (or a crossover wavelength of $\lambda_C \approx 1.7$ cm, below which the effects of the capillary waves become relevant; note this length is different from the capillary length $L_C\approx 2.7$ mm associated to the formation of capillary meniscus). Thus, the range of wave numbers compatible with $\sim k^{-5/2}$ scaling seems to be bounded by these two wavenumbers, $k_F$ and $k_C$.

From the spectra in Fig.~\ref{f:spectra}(c) we can provide another estimation of the amplitude of the turbulent fluctuations, by setting apart the integrated amplitude of the peak of the spectrum $\overline{\eta} = [2\int_{k<60} E(k)dk]^{1/2}$ from the amplitude of the fluctuations $\delta \eta = [2\int_{k \ge 60} E(k)dk]^{1/2}$. Their ratio $\delta \eta / \overline{\eta}$ is close to 5\% in all cases. As a reference, in the turbulent ocean the ratio of turbulent to non-turbulent excitations ranges from 1\% to 10\% \cite{dasaro_turbulence_2014}, while in the turbulent solar wind (an example in which waves and mean flows coexist) the relative amplitude of fluctuations to the mean magnetic field is also close to $10\%$ \cite{bruno2013solar}. Even for strong hydrodynamic turbulence, experiments with turbulent fluctuations of $10\%$ amplitude are not uncommon. Typical experiments of isotropic and homogeneous turbulence in wind tunnels have a ratio of turbulent fluctuations to mean velocities of $10\%$ or less \cite{hurst2007scalings}, while von K\'arm\'an experiments designed to reach very high Reynolds numbers with significant fluctuations can achieve values of this ratio of $\approx 30\%$ \cite{cortet2009normalized}.
 
Once the experiment described above reached a turbulent steady state, the flow was seeded with over 100 floaters (of these, 52 were used on average for the study, as this is the approximate number of floaters observed simultaneously in the central region of the vessel where the measurements were made). The floaters were 3D-printed spherical shells of black polyvinyl chloride (PVC), with a diameter of $15$~mm, a wall thickness of $1$ mm, and a weight of $(0.84 \pm 0.04)$~g (see Fig.~\ref{f:setup}). The particle color was chosen to facilitate their detection, in conjunction with the use of $\textrm{TiO}_{2}$ powder used to dye the water white. The size of the particles was chosen of the order of the crossover wavelength $\lambda_C$, in such a way that particle motions were affected mostly by gravity waves (which have longer wavelengths). Particles were designed to have good buoyancy, and low interaction with each other and with the fluid. The specific gravity of the particles was $\overline{\rho_p}/\rho_w \approx 0.5$ (where $\overline{\rho_p}$ is the mean density of the particles and $\rho_w$ is the density of water at 4 $^\circ$C), and as a result half of each particle was submerged while the other half was above the surface. Also, note that even though Archimedes' principle ensures that the specific gravity computed over only the half of the particle that is submerged is equal to one, capillary effects can also make the floaters inertial (i.e., in practice, heavier or lighter than the displaced water) \cite{denissenko2006waves}.

Particle tracking was done using the PTV technique. Images were captured with a $120$~Hz acquisition frequency using two ultra-fast cameras (a 1024 PCI Photron camera and a SA3 Photron camera). Both cameras were synchronized using an external trigger, and the experiment was illuminated using two high-power halogen lamps with low flicker. 3D particle trajectories were then reconstructed from the images using a computer cluster, from which particle velocities and accelerations could also be computed. Each experimental run consisted of 23~s of acquisition time, and a total of 7 experiments were performed for each maximum forcing amplitude $A$ explored (resulting in a total of 28 experiments considered for the present study).

\begin{figure}
\centering
\includegraphics[width=8.5cm]{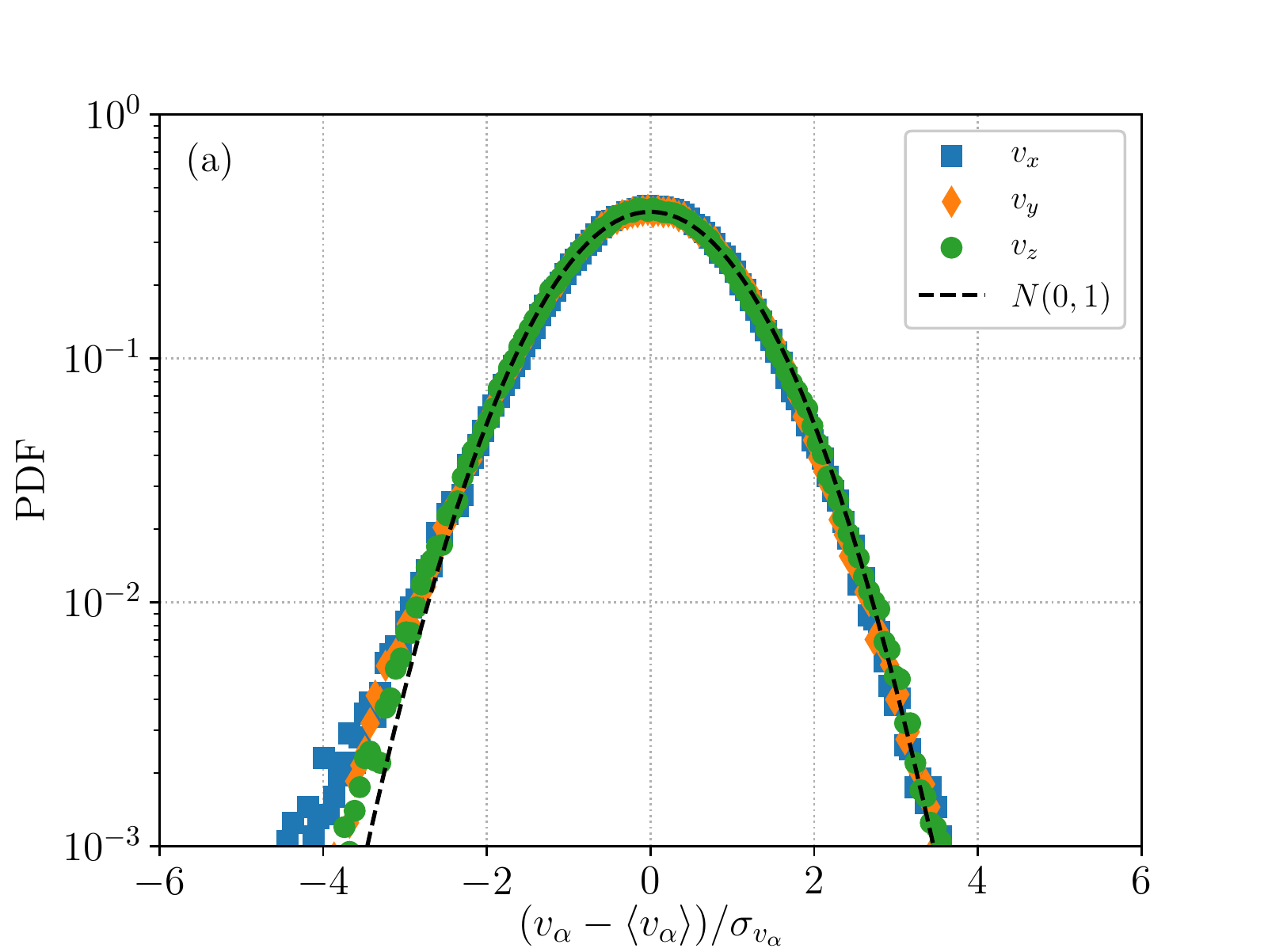}
\includegraphics[width=8.5cm]{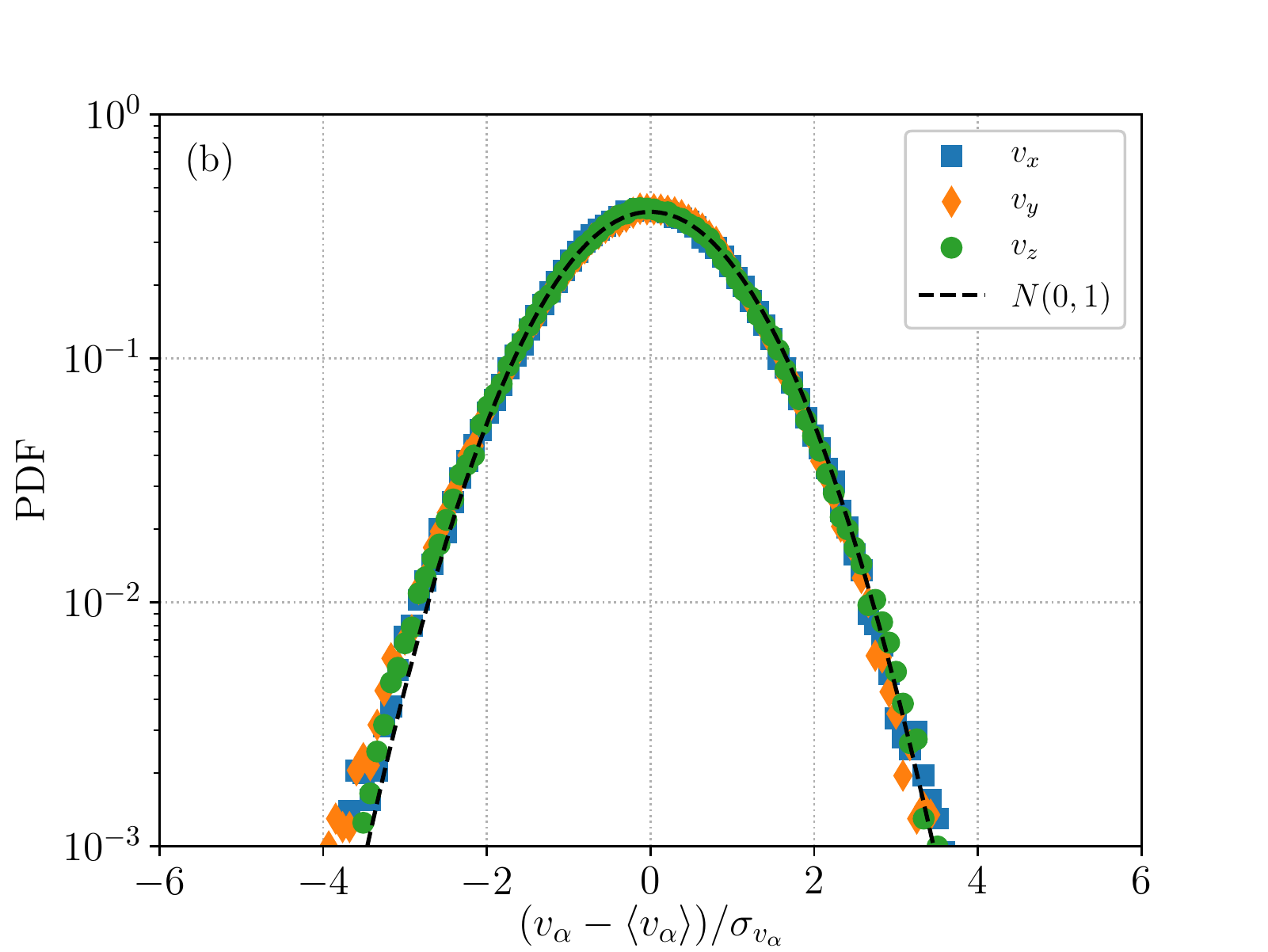} \\
\includegraphics[width=8.5cm]{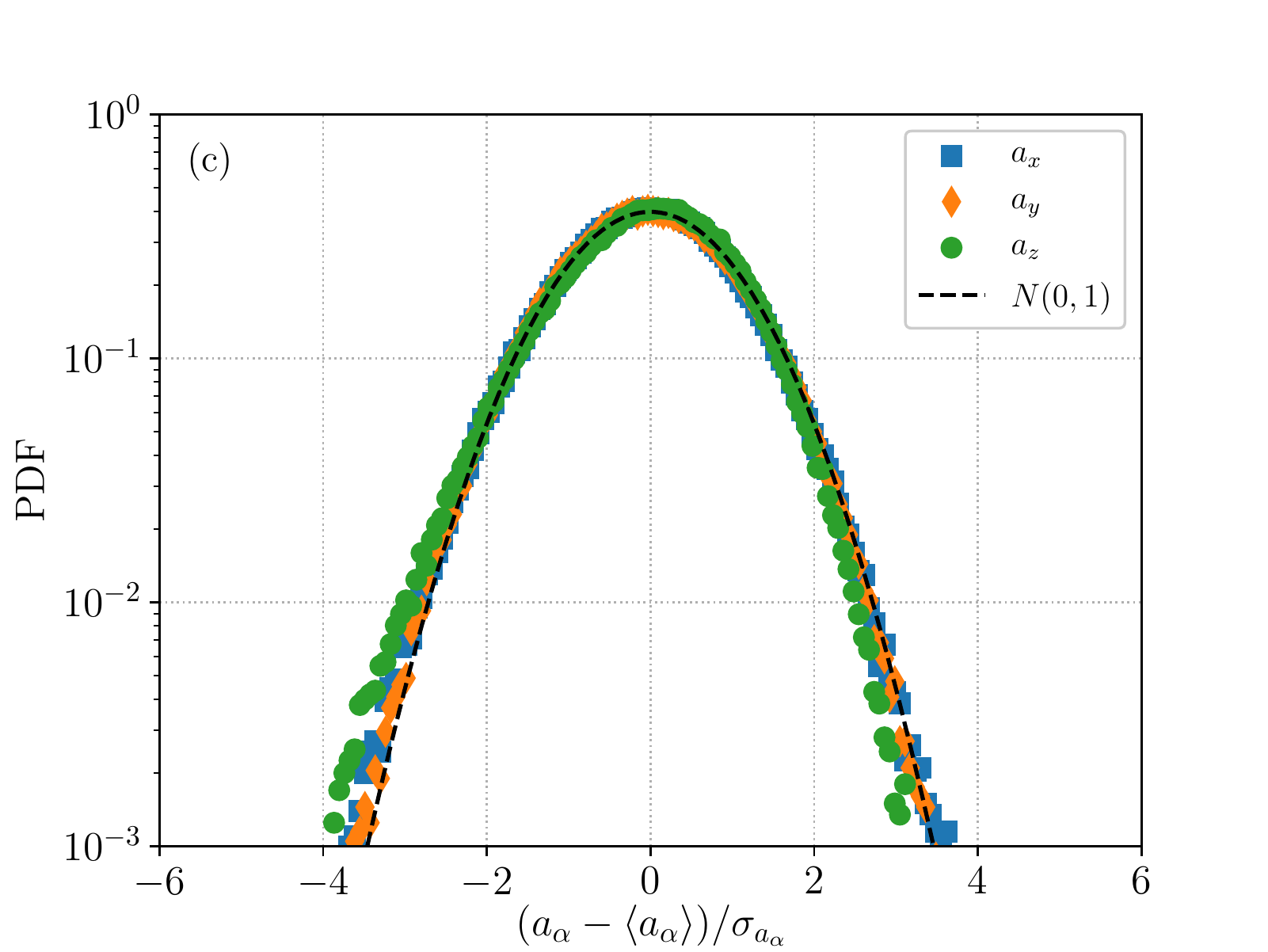}
\includegraphics[width=8.5cm]{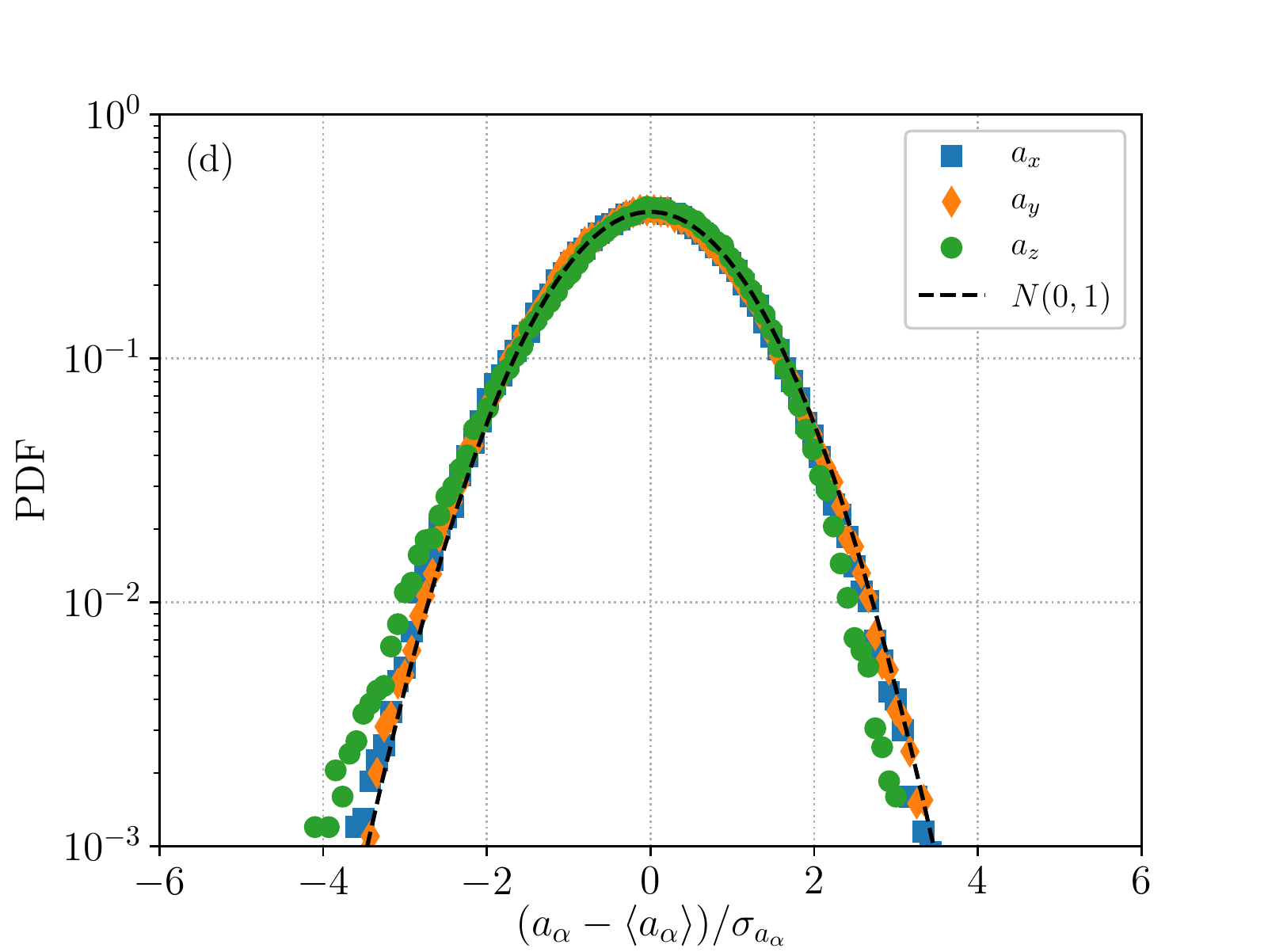}
\caption{({\it Color online}) {\it Top}: PDFs of the Cartesian components of the fluctuating velocity of the floaters for (a) experiment A10, with maximum spatial displacement of the paddles of $A=10$ mm, and (b) for experiment A20, with maximum spatial displacement of the paddles of $A=20$ mm. Velocities are normalized by their dispersions, and thus are dimensionless. {\it Bottom}: PDFs of the Cartesian components of the fluctuating acceleration of the floaters in the same experiments, for (c) $A=10$ mm, and (d) $A=20$ mm. As for the velocity, the accelerations are normalized by their corresponding dispersions. Labels for the three Cartesian components are given in the insets. In all cases, we also show as a reference a normal distribution centered around zero with dispersion of unity, represented by the solid black curve and denoted as $N(0,1)$.}
\label{f:A_V}
\end{figure}

\section{Velocity and acceleration \label{sec:vel}}

We start characterizing particle velocities and accelerations. As will be seen later, significant information on the flows in the experiments can be obtained from the PTV measurements. Figure \ref{f:A_V} shows the probability density functions (PDFs) of the three Cartesian components of the floaters' velocity and acceleration, for experiments A10 and A20 (respectively with maximum spatial displacement of the paddles of $A=10$ mm and $A=20$ mm, see table ~\ref{tab:vel_accel_vs_forcing}). For the PDFs in Fig.~\ref{f:A_V}, the mean velocities and accelerations were subtracted, and the results were normalized by their corresponding dispersions. 

\begin{table}
\centering
\begin{ruledtabular}
\begin{tabular}{l c c c c c c c c c c c c c c c c c c c c}
Experiment & 
$A$ [mm] & 
$\langle | v_x | \rangle$ &
$\langle | v_y | \rangle$ &
$\langle | v_z | \rangle$ &
$\langle | a_x | \rangle$ &
$\langle | a_y | \rangle$ &
$\langle | a_z | \rangle$ &
$\sigma_{v_x}$ &
$\sigma_{v_y}$ &
$\sigma_{v_z}$ &
$\sigma_{a_x}$ &
$\sigma_{a_y}$ &
$\sigma_{a_z}$ \\
\hline
A05 & 5 & 24.28 & 20.36 & 38.78 & 444.1 & 373.7 & \, 766.5 & 30.65 & 25.65 & 48.84 & 588.4 & 471.7 & \, 965.8  \\ 
A10 & 10 & 38.56 & 37.19 & 60.56 & 660.4 & 657.1 & 1179.7 & 49.20 & 46.84 & 76.60 & 834.1 & 824.6 & 1486.7 \\
A15 & 15 & 44.05 & 44.29 & 67.81 & 718.5 & 755.8 & 1297.3 & 55.80 & 55.99 & 85.32 & 906.7 & 950.3 & 1624.5  \\
A20 & 20 & 48.48 & 48.33 & 71.16 & 748.6 & 787.1 & 1327.5 & 61.25 & 61.06 & 89.91 & 941.5 & 987.1 & 1666.5 \\
\end{tabular}
\end{ruledtabular}
\caption{Labels, parameters, and characteristic values for all experiments. The maximum forcing amplitude $A$ is listed, followed by the mean absolute value of each velocity ($v_\alpha$, for $\alpha=x$, $y$, or $z$) and acceleration ($a_\alpha$) component, and the corresponding standard deviation of each component of the particles' velocity ($\sigma_{v_\alpha}$) and acceleration ($\sigma_{a_\alpha}$). The forcing amplitude increases linearly from the top row to the bottom row of the table. The values or $v_\alpha$ and $\sigma_{v_\alpha}$ are expressed in units of mm s$^{-1}$, while those for $a_\alpha$ and $\sigma_{a_\alpha}$ are given in units of mm s$^{-2}$.}
\label{tab:vel_accel_vs_forcing}
\end{table}

In table~\ref{tab:vel_accel_vs_forcing} we also provide the mean absolute values and standard deviation of each component of the particles' velocity and acceleration. Each ``experiment'' is labeled by its maximum amplitude $A$ (note that, as already mentioned, each one of them actually corresponds to an ensemble of 7 experiments). As the forcing amplitude is increased linearly (from top to bottom of the table), all quantities in table~\ref{tab:vel_accel_vs_forcing} show a monotonic (nonlinear) increase with $A$. The growth in the mean absolute horizontal velocities and in their dispersions as $A$ is increased is associated with two physical effects that become evident from visual inspection of the particles' trajectories, and also from the discussions that follow: as $A$ is increased, mid-size horizontal eddies develop in the flow, as well as a large-scale circulation in the entire vessel that advects the particles in the $x$-$y$ plane with a well defined mean velocity. However, as we will see later, this large-scale circulation in the vessel has a small velocity, and thus the main contribution to the observed increase in horizontal velocities as $A$ is increased is related to drag of the particles by the waves and to advection by mid-size eddies. The vertical velocity (as well as its dispersion) also increases with increasing $A$, as also do all Cartesian components of the acceleration and their corresponding dispersions.

After subtracting these mean components, the PDFs of the fluctuating velocity in Fig.~\ref{f:A_V} are close to a normal distribution, although a small asymmetry can be seen in the PDFs of $v_x$ and $v_y$ towards negative values of the velocity. PDFs of the acceleration components are also close to Gaussian, although a weak asymmetry can now be seen instead in the vertical acceleration. As a result, values of the skewness and kurtosis for these quantities are close to normal values (skewness of the acceleration is between $\approx -0.3$ and 0, and of the velocity is between $\approx -0.2$ and 0, while the kurtosis of the acceleration and velocity components varies between $3$ and $3.4$ with no clear dependence on the forcing amplitude $A$).

\begin{figure}
\centering
\includegraphics[width=8.5cm]{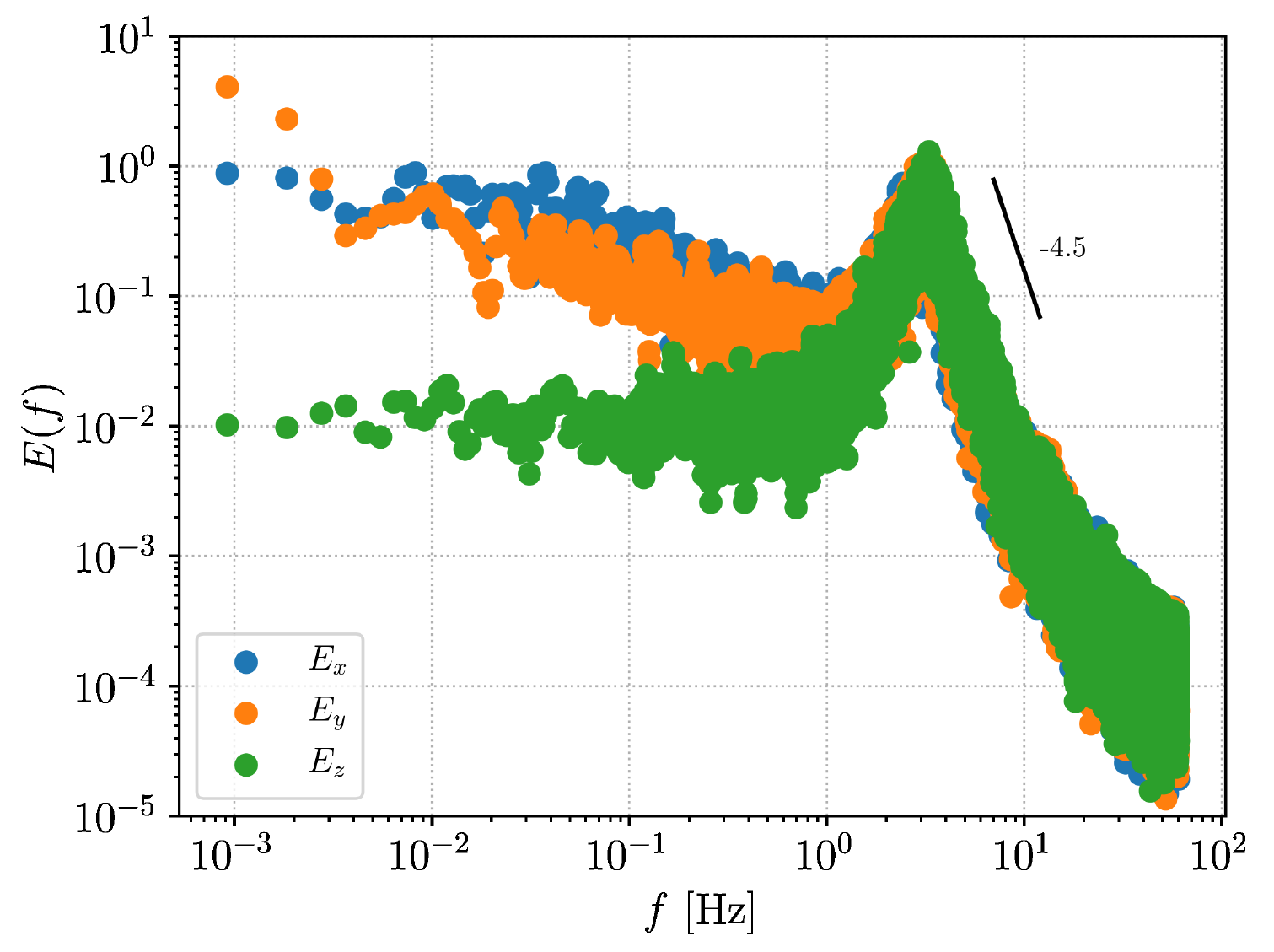}
\includegraphics[width=8.5cm]{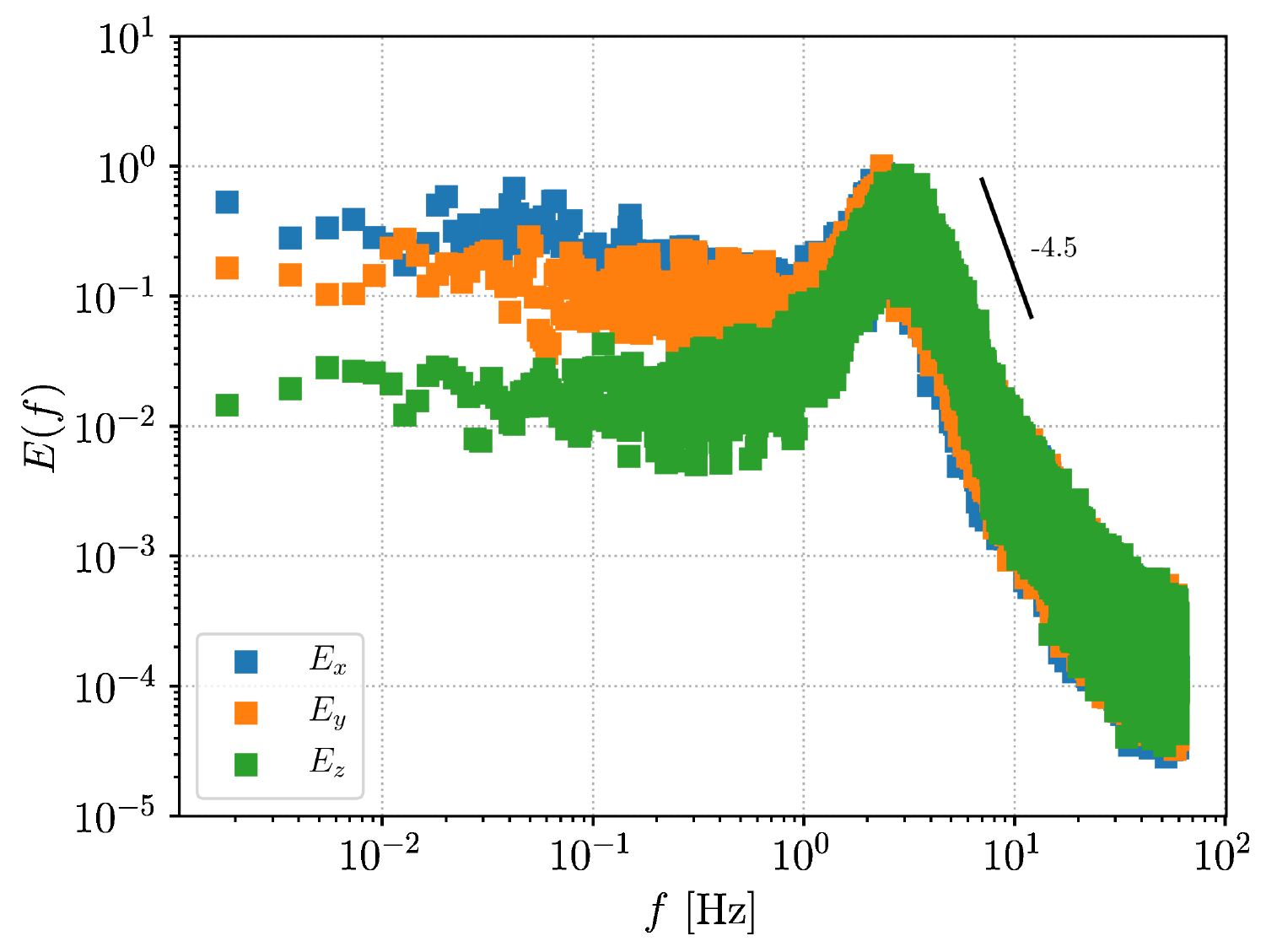}
\caption{({\it Color online}) Kinetic energy spectra of the floaters, for each Cartesian component of the velocity, for experiments A10 ({\it left}) and A20 ({\it right}). A power law $E(f) \sim f^{-4.5}$ for frequencies larger than 4 Hz is shown as a reference.}
\label{f:Ef}
\end{figure}

As the floaters move with the free surface in the vertical direction, and are advected and dragged by the fluid in the horizontal direction, we can use the measured velocity components to obtain horizontal kinetic energy spectra of the particles $E_x(f)$ and $E_y(f)$ (corresponding respectively to the power spectra of the particles velocities $v_x$ and $v_y$), and a vertical kinetic energy spectrum $E_z(f)$ (obtained from $v_z$, which, as the particles move together with the surface, provides an estimation of the surface Lagrangian vertical velocity spectrum). The resulting spectra are shown in Fig.~\ref{f:Ef}, for experiments A10 and A20 (left and right panels, respectively). There are only minor differences in the spectra as $A$ is varied. In both cases shown, horizontal spectra display a power law for small frequencies, a peak near the forcing frequency at $\approx 3$ Hz, and then a fast power law decay for higher frequencies  compatible with a $\sim f^{-4.5}$ decay. The vertical spectrum also displays the peak at $\approx 4$ Hz, followed by the same power law at higher frequencies, but for low frequencies the spectrum is flat. Interestingly, these spectra are qualitatively similar to those found in oceanic observations using buoys, such as, e.g., those from $1$ m buoys in the Pacific Northwest \cite{dasaro_lagrangian_2000}.

\begin{figure}
\centering
\includegraphics[width=8.6cm]{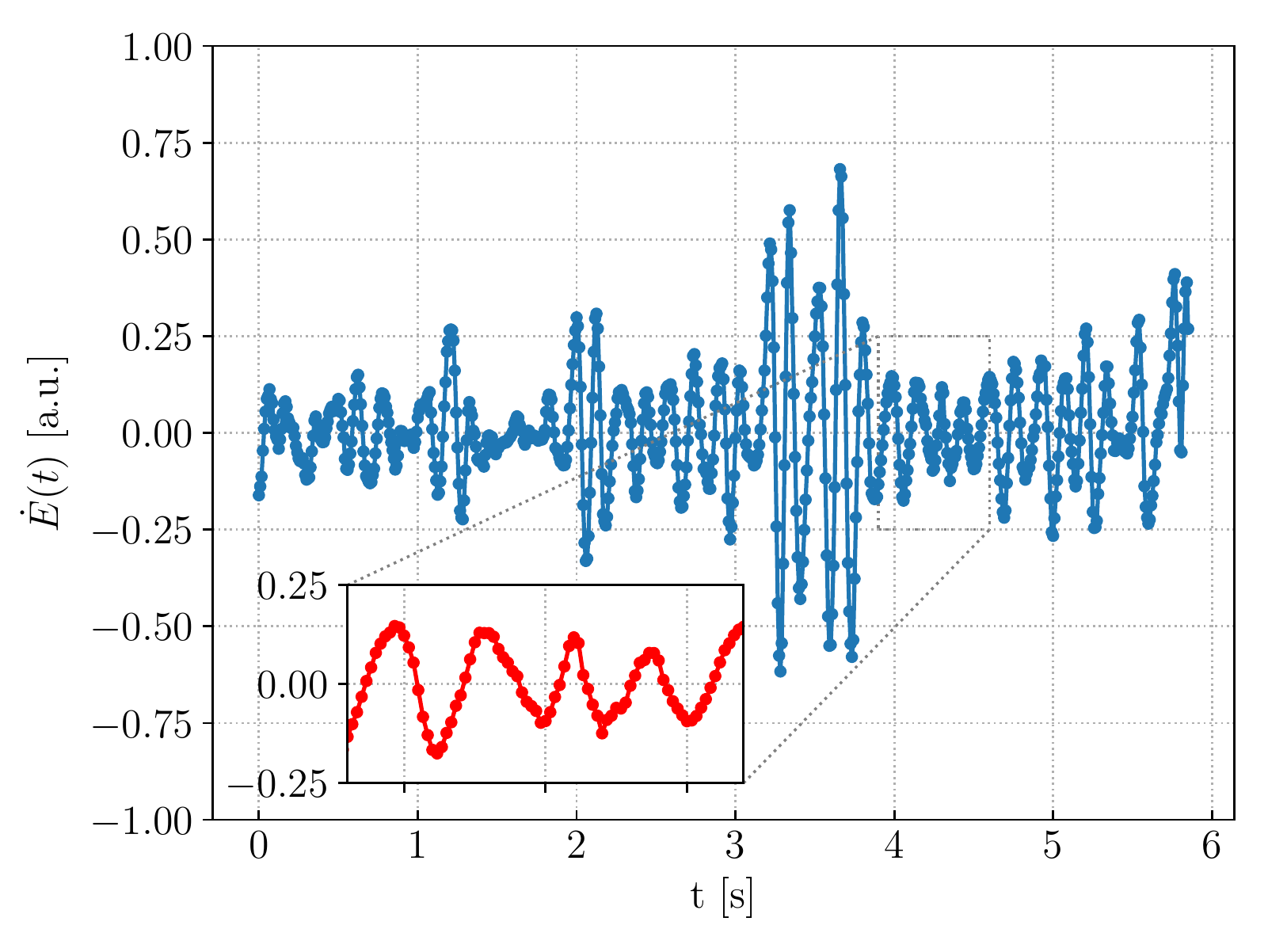}
\includegraphics[width=8.3cm]{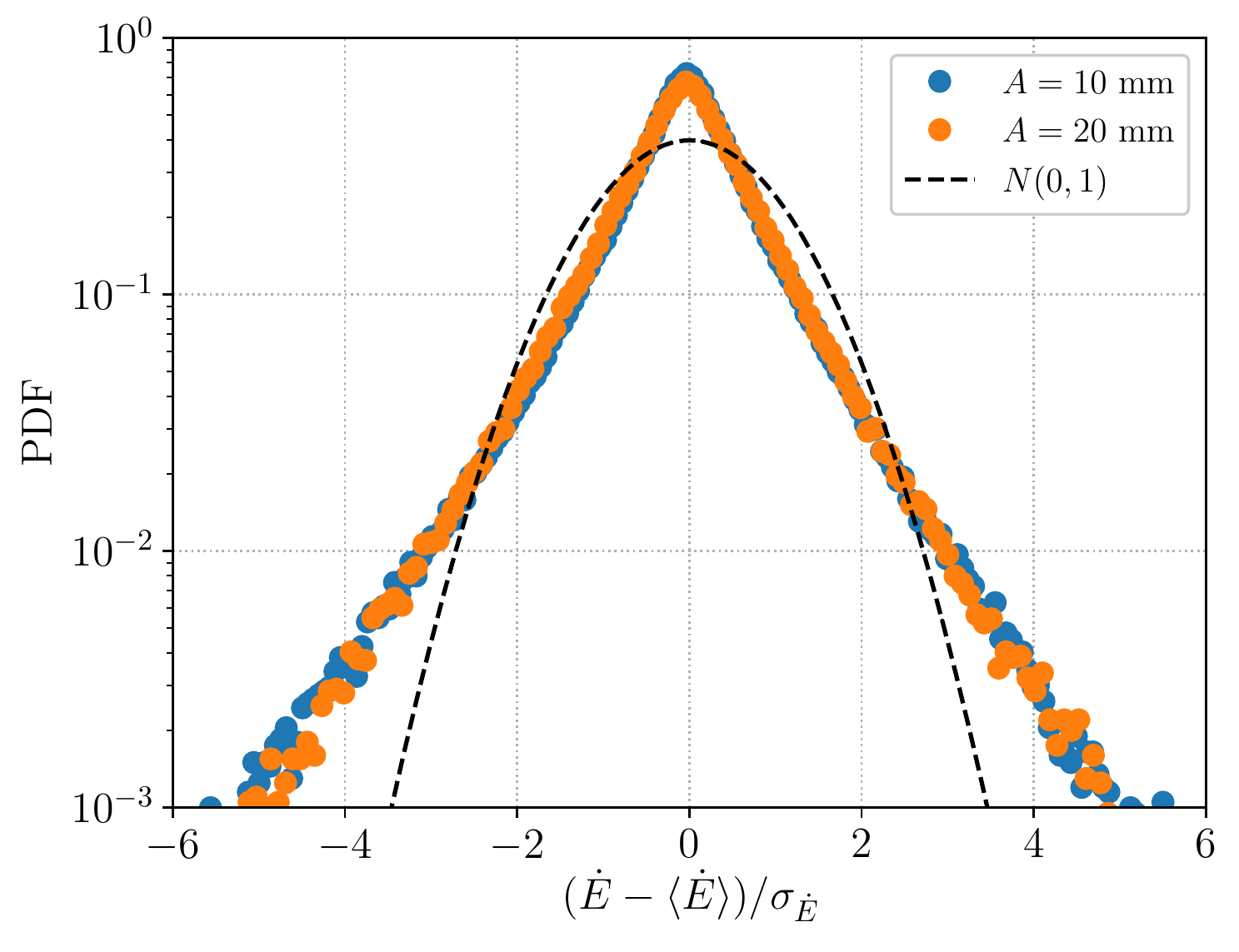}
\caption{({\it Color online}) {\it Left:} Instantaneous power $dE/dt$ gained or lost by a particle in experiments A10 (with maximum forcing amplitude $A=10$~mm). The inset shows a detail of the time series. {\it Right:} PDF of the instantaneous power gained or lost by the particles, in experiments A10 and A20. The shape of the PDF is the same for both experiments. For completeness, it is worth mentioning that the $\sigma_{\dot{E}}$ value associated with the two experiments is $\sigma_{\dot{E}}^{\text{A10}} = 0.14$~mW and $\sigma_{\dot{E}}^{\text{A20}} =0.18$~mW, respectively. A normal distribution is shown as a reference by the black solid curve, and labeled as $N(0,1)$.}
\label{f:Et}
\end{figure}

Finally, in Fig.~\ref{f:Et} we show a time history of the power gained or lost by a particle, $dE/dt$, and its PDF for all particles in experiment A10. To reduce numerical errors, the power $dE/dt$ was estimated as the dot product between the particles speed and acceleration. The PDF is symmetric but with strong tails, deviating significantly from a normal distribution, indicating that particles can gain or loose energy regardless of their velocities, in an intermittent fashion. However, note the non-Gaussianity of $dE/dt$ can simply arise from the fact that the power gained or lost by the particles is a non-linear function of Gaussian quantities. The time series of the instantaneous power is compatible with the observed probability distribution, as short periods of time in which $dE/dt$ reaches large values can be seen in the figure.

\section{Particle trajectories and dispersion\label{sec:part}}

\subsection{Experimental results}

\begin{figure}
\begin{center}$
\begin{array}{rr}
\text{\bf{(a)}}
\includegraphics[width=9.5cm]{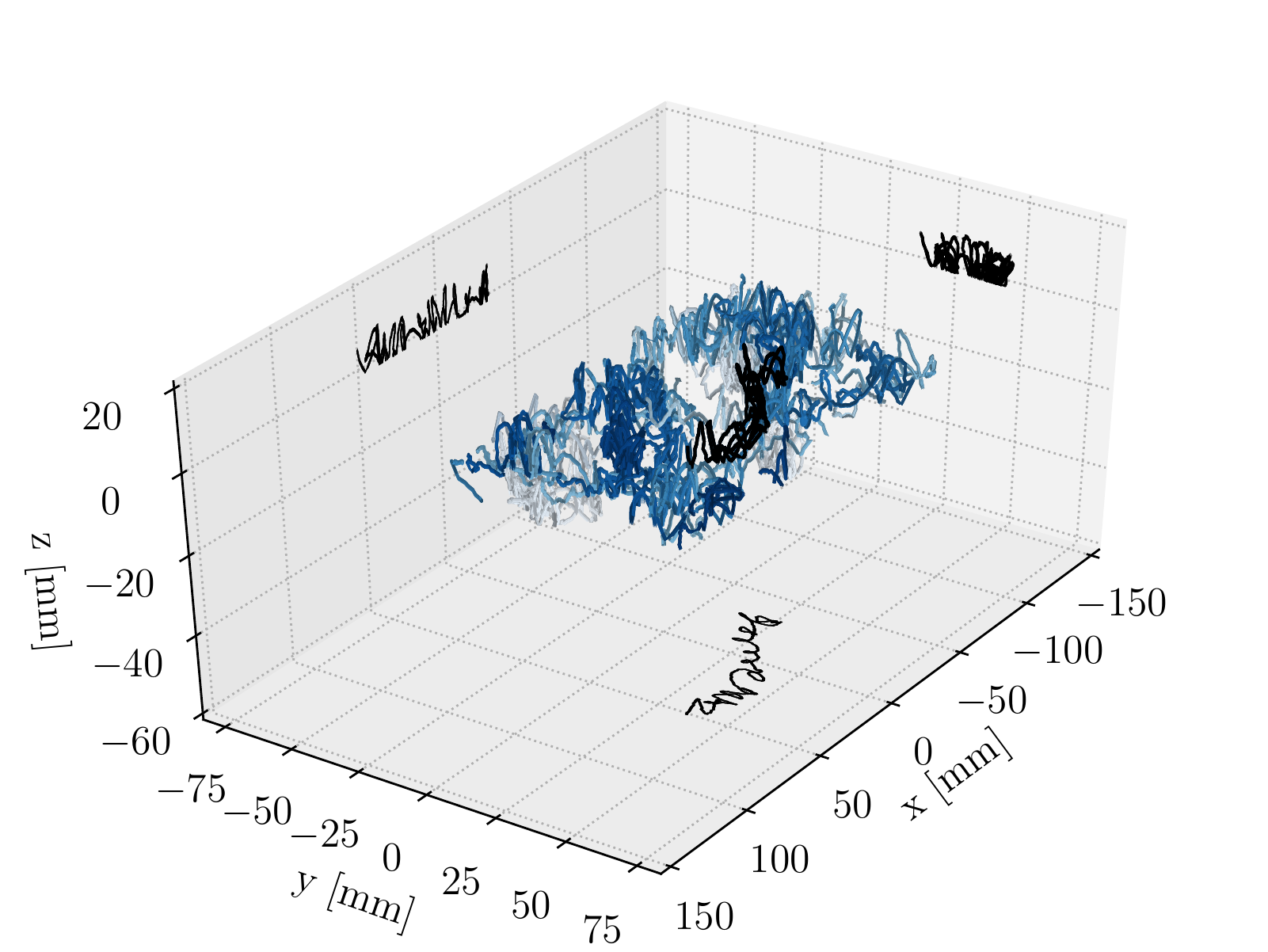} &
\end{array}
\begin{array}{rrr}
\text{\bf{(b)}}
\includegraphics[width=6.4cm]{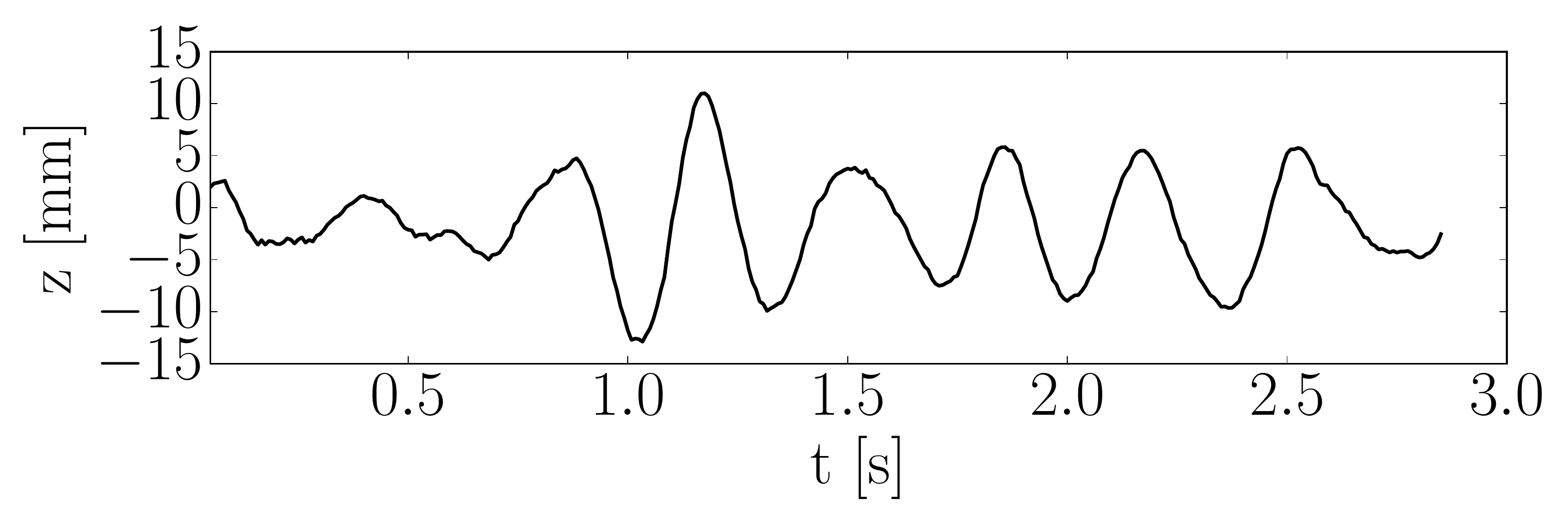} \\
\text{\bf{(c)}}
\includegraphics[width=6.5cm]{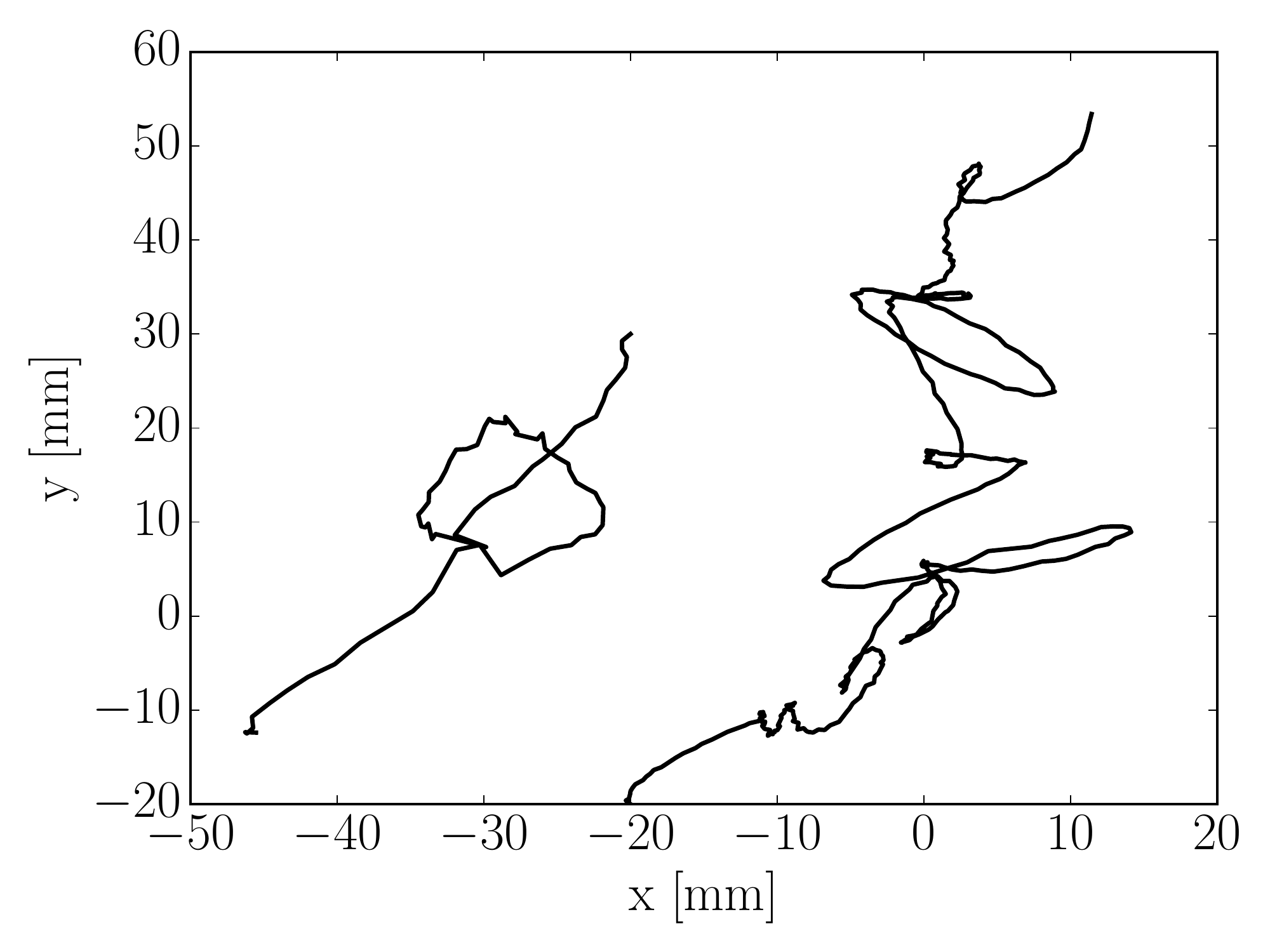} 
\end{array}$
\end{center}
\caption{({\it Color online}) (a) Three-dimensional trajectories of 52 particles in experiment A10, in light gray and blue. The trajectory of one particle is highlighted in black, and projected into the $x$-$y$, $x$-$z$, and $y$-$z$ planes. Note, in the projection onto the $x$-$y$ plane, the coexistence of three phenomena: fast oscillations associated to waves, circular trajectories associated to eddy trapping, and a slow drift. (b) Vertical position $z$ of a particle as a function of time in the same experiment. All other particles follow a similar vertical behavior. (c) Horizontal trajectories of two particles in the same experiment. Note the circular motions in the particles trajectories.}
\label{f:xyz}
\end{figure}

\begin{figure}
\includegraphics[width=8.5cm]{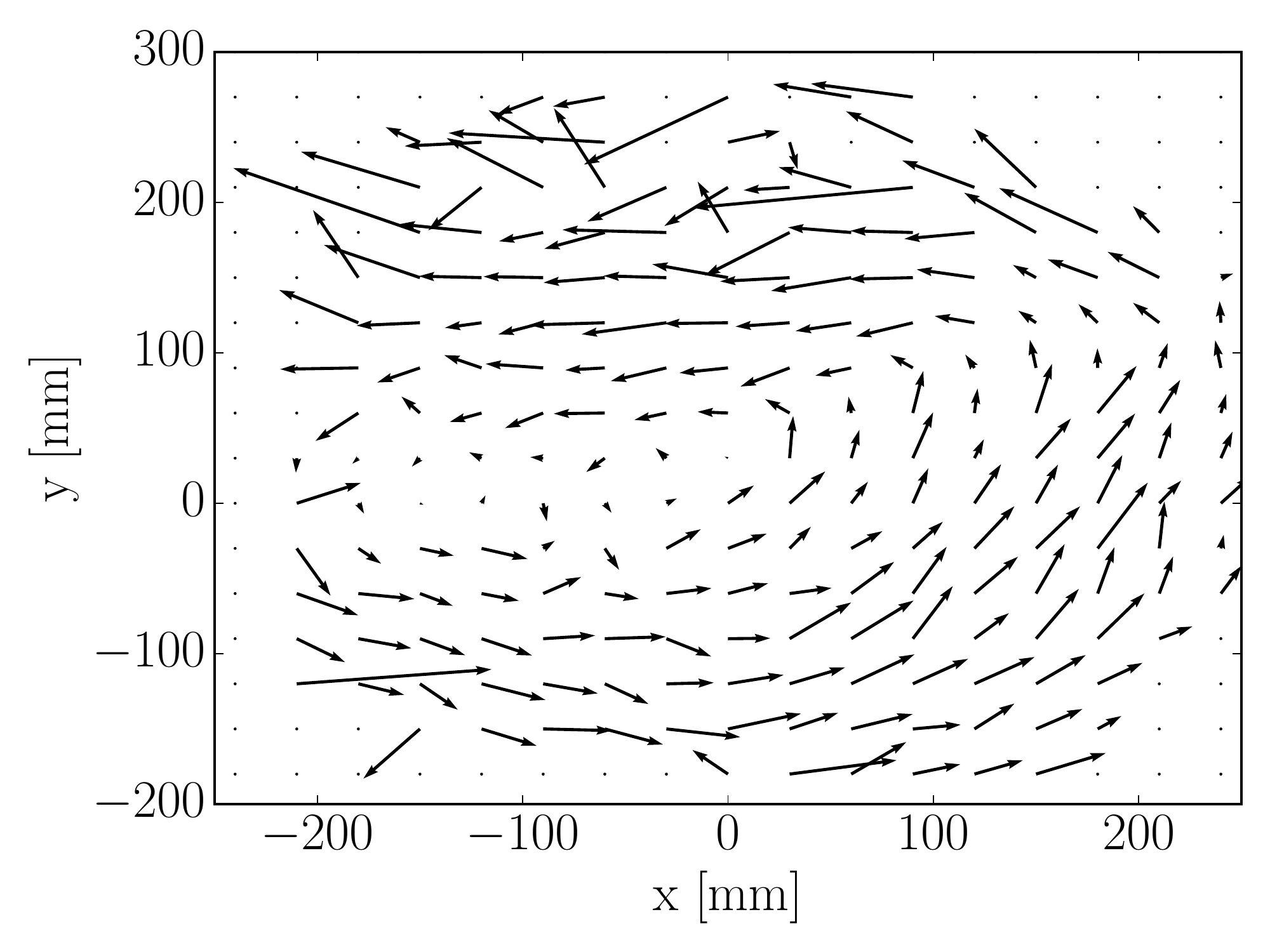}
\caption{Temporal average, in an Eulerian grid, of the horizontal velocities of all particles in the observed region of experiment A10.  Note the mean large-scale circulation that develops in the experiment.}
\label{f:circulation}
\end{figure}

\begin{figure}
\centering
\includegraphics[width=8.5cm]{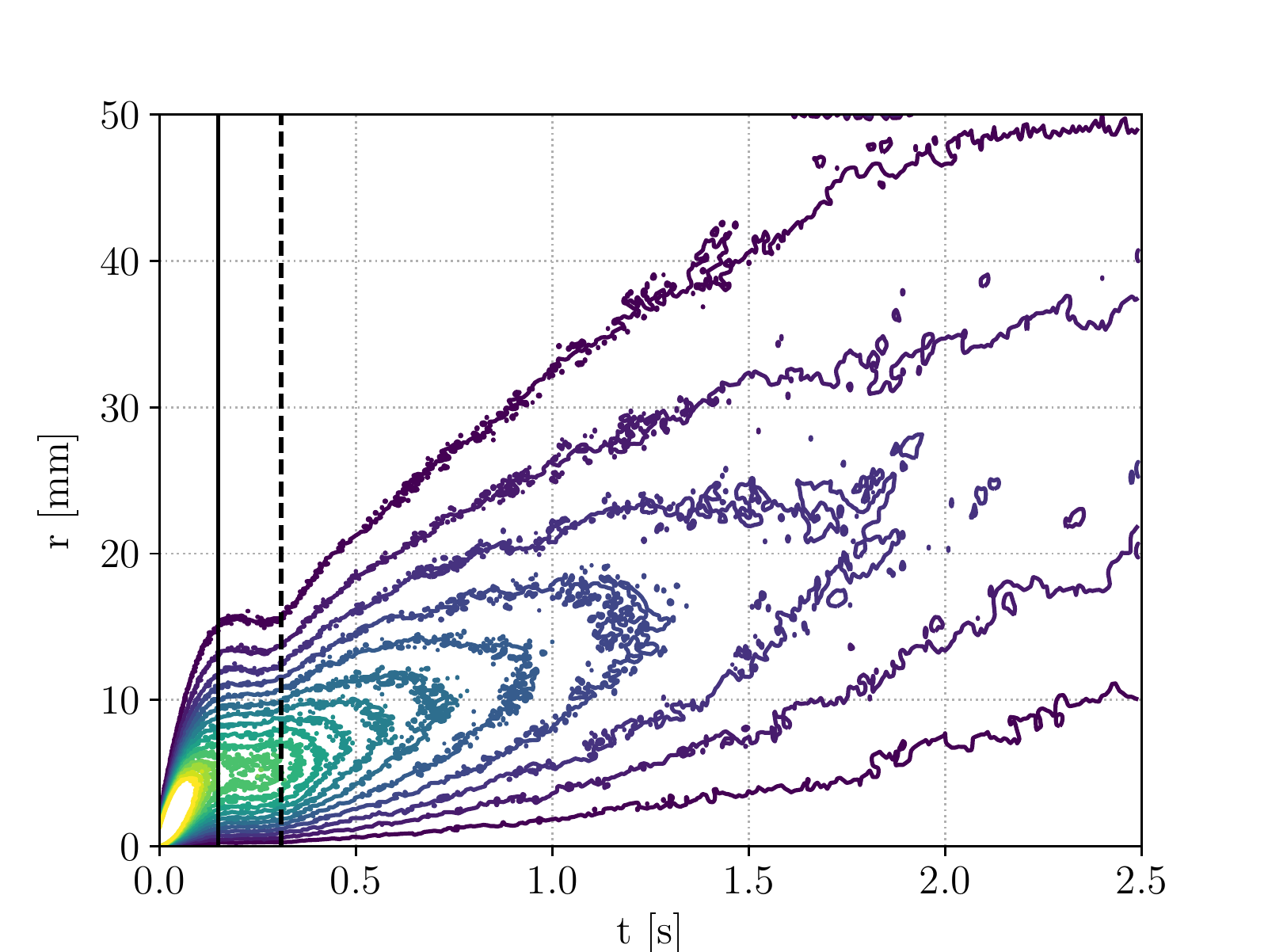}
\includegraphics[width=8.5cm]{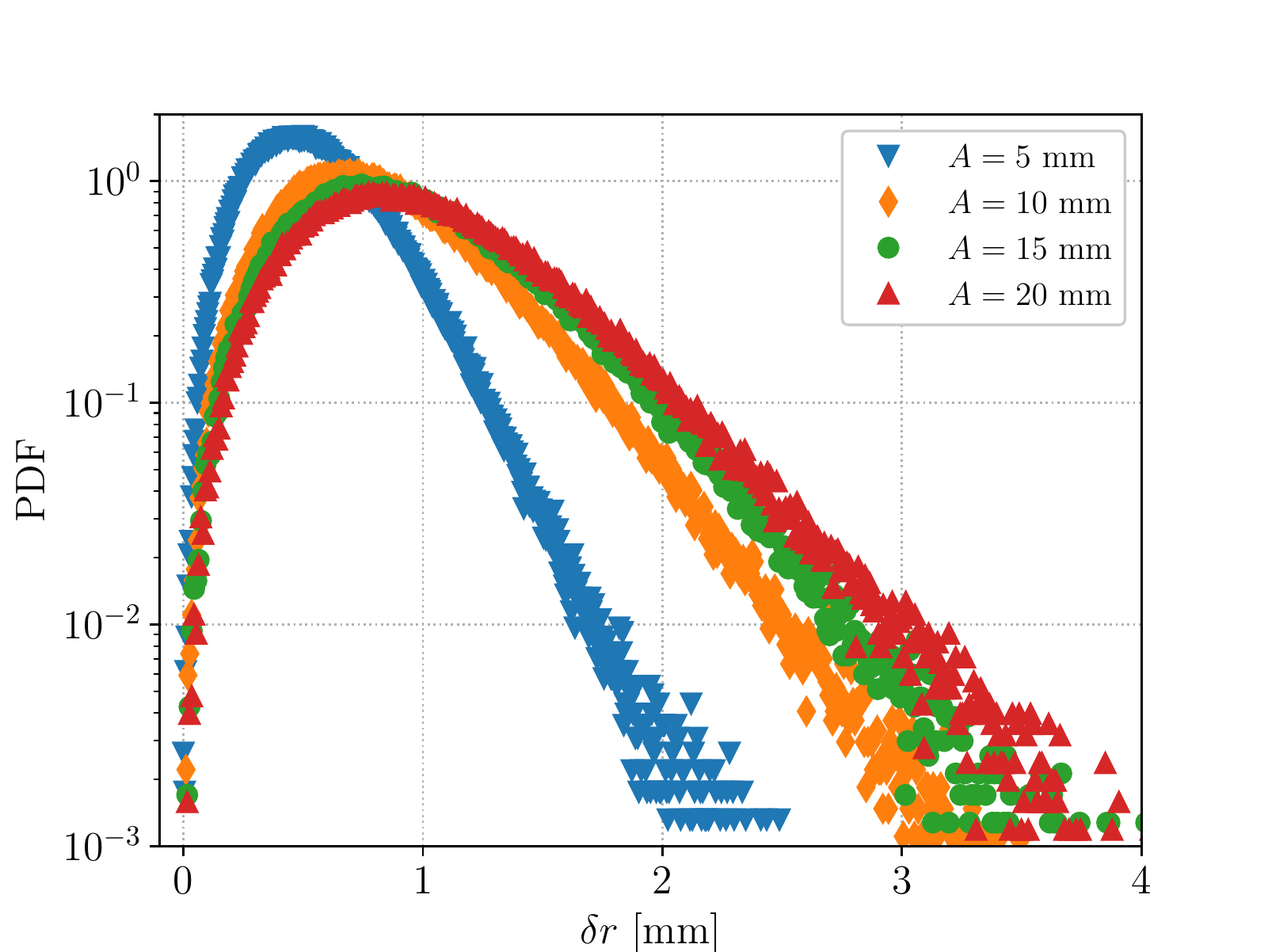}
\caption{({\it Color online}) {\it Left}: Isocontours of the joint probability distribution $P(r,t)$ of particles being displaced a horizontal distance $r=\delta r$ at a time $t$, in experiment A10. The two vertical lines separate three different regimes as explained in the text. {\it Right}: PDFs of the horizontal displacements $P(r,t)$ at a fixed time and as a function of the displacement $r$, for all the experiments.}
\label{f:dr}
\end{figure}

As already mentioned, in each realization of the experiments we followed the three-dimensional trajectories of approximately $52$ particles. Figure \ref{f:xyz}(a) shows as an illustration the resulting trajectories for just one realization of an experiment. The trajectory of one individual particle is highlighted, and projected into the horizontal and vertical planes. For clarity, and to distinguish some typical features of the particles' trajectories, the evolution of another particle is shown in Fig.~\ref{f:xyz}(b) only for its vertical displacement $z(t)$ as a function of time, and in Fig.~\ref{f:xyz}(c) for two particles at different times in the $x$-$y$ plane (all from the same realization of the experiment). The behavior in Fig.~\ref{f:xyz}(b) is typical of all particles, a wave-like motion (with multiple time scales) can be clearly seen in the vertical direction. Instead, in the $x$-$y$ plane we can differentiate multiple features. Particles move back and forth in a fast time scale, associated with the waves. But particles also follow circular trajectories, which can be associated with transient trapping of the particles by horizontal eddies in the flow (other ``trapping'' events can be caused by superposition of waves traveling in different directions or by nodal regions of quasi-standing waves, although as we will show below, all these events can be modeled simply by considering the effect of the eddies). Finally, particles in Fig.~\ref{f:xyz}(c) can be also seen to slowly drift in one direction. This mean displacement is generated by the large-scale circulation that develops in the vessel. To illustrate this effect, Fig.~\ref{f:circulation} shows a mean horizontal (Eulerian) velocity field for the observed region in experiment A10, reconstructed by interpolating the particles' velocities into a fixed array of points and by computing an average in time. The presence of a global circulation can be observed. Thus, three phenomena can be identified from the particles trajectories: displacement of the particle by waves in the vertical direction (as well as a fast modulation in the same time scales in the $x$-$y$ plane), circular trajectories associated to the presence of horizontal eddies in the $x$-$y$ plane, and a slow and coherent drift in the same plane. Indeed, from visual inspection it can be seen that as $A$ is increased, the paddles not only excite surface waves, but also excite mid-size horizontal eddies that detach from the side of the paddles, as well as the slow circulation with a characteristic length of the size of the vessel and that slowly builds in time as the experiment progresses.

With all these trajectories we can compute the single-particle vertical and horizontal displacement, respectively defined for each particle as
\begin{equation}
\delta z_i = z_i(t) - z_i(0) , \,\,\,\,\,\,\,\, \delta r_i = r_i = \sqrt{[x_i(t) - x_i(0)]^2 + [y_i(t) - y_i(0)]^2} ,
\end{equation}
where the subindex $i$ labels the particle. Figure \ref{f:dr} shows the resulting joint probability distribution $P(r,t)$, of finding a particle being displaced a horizontal distance $r$ at a time $t$, again for experiment A10. As can be seen from the figure, the maximum of the PDF increases with time, which can be expected: as time evolves the most probable horizontal displacement $r$ increases, as floaters are dispersed by the turbulence. However, there are three distinct regimes (separated by the vertical lines in the figure): at early times the most probable value of $r$ increases with time, then it saturates, and finally it increases again. Figure \ref{f:dr} also shows $P(r,t)$ for the four experiment sets at a fixed time and as a function of $r$. The observed PDFs are close to a Rayleigh distribution. This is the expected distribution for particles performing a random walk in the horizontal plane, and compatible with a diffusion process in the free surface of the fluid. In addition, note that as the amplitude of the forcing $A$ is increased, the most probable displacement (i.e., the position of the peak of the PDF) increases. However, note also that there is a strong dispersion of $r$ around these values.

\begin{figure}
\centering
\includegraphics[width=8.5cm]{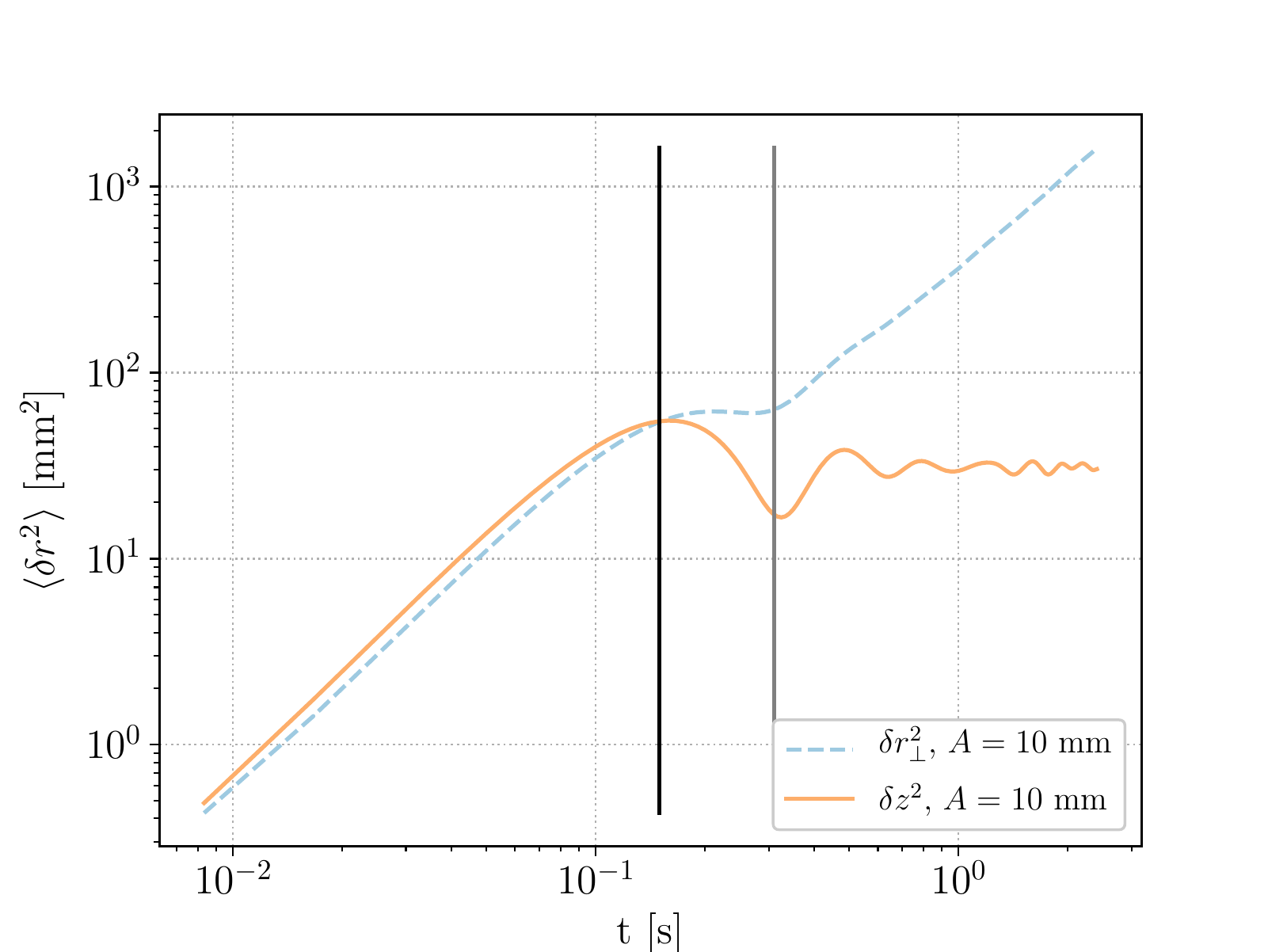}
\includegraphics[width=8.5cm]{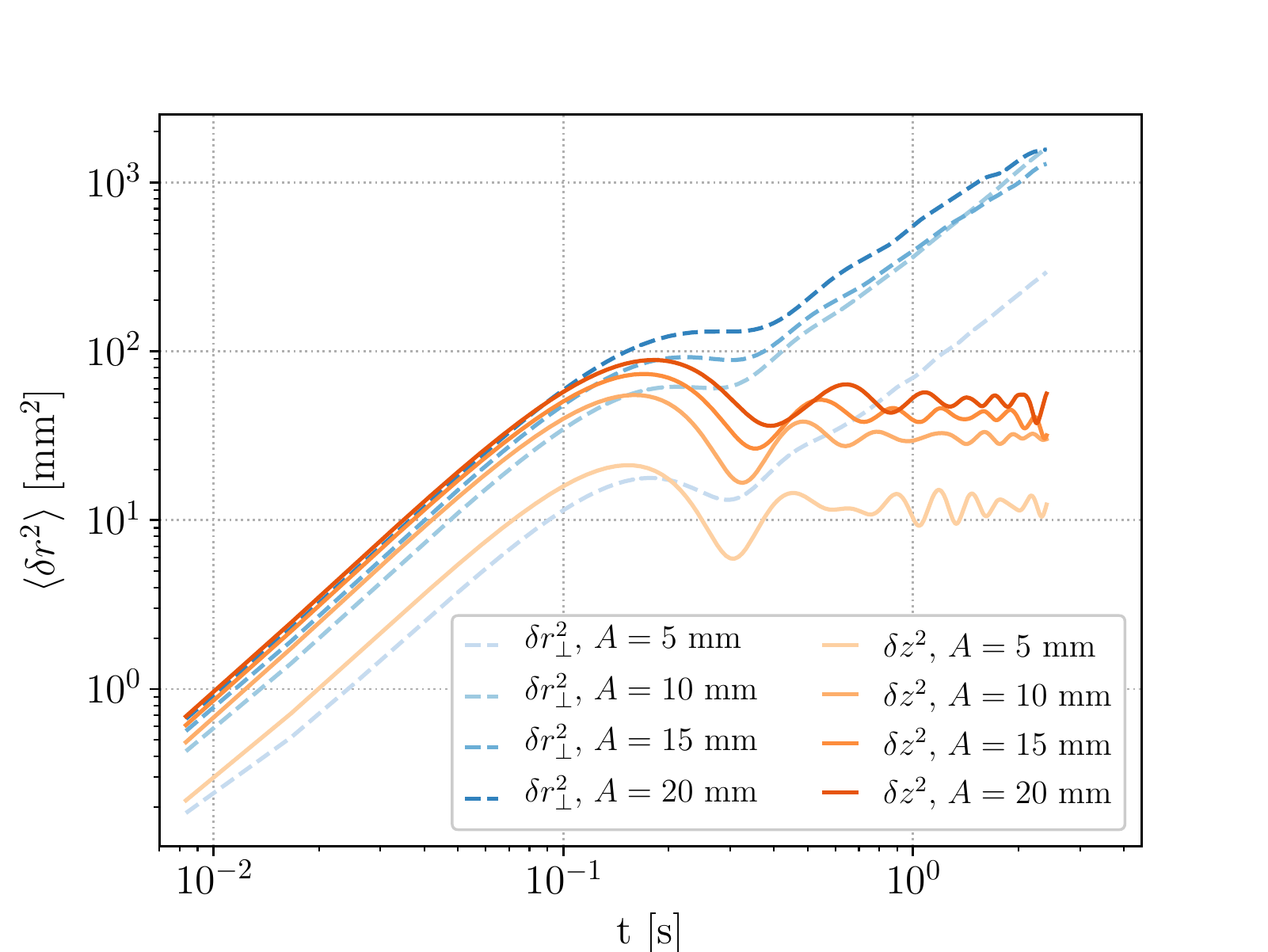}
\caption{({\it Color online}) {\it Left:} Mean quadratic vertical and horizontal displacement of the floaters in experiment A10, as indicated in the inset. Note the three regimes, separated as a reference by the two vertical lines. These lines indicate, from left to right, the inverse of half the frequency of the most energetic waves, and the particle velocity correlation time. {\it Right:} Mean quadratic vertical and horizontal displacement of the floaters in all experiments.}
\label{f:DCM}
\end{figure}

The mean quadratic displacements can be obtained from the PDFs or as the second order moment of the single-particle vertical and horizontal displacements,
\begin{equation}
\left< \delta z^2 \right>(t) = \left<[z_i(t) - z_i(0)]^2 \right> , \,\,\,\,\,\,\,\, \left<\delta r^2 \right>(t) = \left<[x_i(t) - x_i(0)]^2 + [y_i(t) - y_i(0)]^2\right> ,
\end{equation}
where the average is performed over all particles. Figure \ref{f:DCM} shows the resulting horizontal and vertical quadratic displacements, first for experiment A10 and then for all experiments. Let's consider first the experiment with fixed forcing amplitude $A$. The mean quadratic vertical dispersion has a ballistic behavior up to a characteristic time of half of the inverse of the fundamental frequency $\approx 3$ Hz. This behavior corresponds to the particles being displaced vertically by the most energetic waves, for half a wave period (i.e., until the maximum in the amplitude of the waves is reached). Then, the ballistic behavior in $\langle \delta z^2 \rangle$ is followed by a saturation towards a mean stationary value for longer times. This can be expected: at the beginning floaters are quickly displaced vertically by the surface waves, but once the slowest wave reaches its maximum, floaters can only oscillate vertically around the equilibrium position. Thus, the value of $\left< \delta z^2 \right>$ for long times is just proportional to the mean squared amplitude of the waves. In the horizontal direction the behavior is more interesting: at early times the mean squared horizontal displacement also grows ballistically, $\left<\delta r^2 \right> \sim t^2$, and also saturates at the first characteristic time. Then $\left<\delta r^2 \right>$ remains approximately constant or grows very slowly for a short period of time, and at a later time it starts to grow again but slightly slower than ballistically in time ($\left<\delta r^2 \right> \sim t^{1.6}$ from a best fit adjustment, an exponent close to the arithmetic average of the random walk and ballistic exponents). Note that the time evolution of $\left<\delta r^2 \right>$ is just the second order moment of the joint probability $P(r,t)$ shown in Fig.~\ref{f:dr}, and thus also follows the same dynamics. When looking at all the experiments (also shown in Fig.~\ref{f:DCM}) we see that the overall shapes of $\left< \delta z^2 \right>(t)$ and of $\left< \delta r^2 \right>(t)$ do not change as we vary the amplitude of the forcing $A$, but that the amplitudes of $\left< \delta z^2 \right>$ and of $\left< \delta r^2 \right>$ are sensitive to $A$, increasing as $A$ is increased.

The mean squared vertical displacements at all times, and the mean squared horizontal displacements at early times, are compatible with the dispersion of particles expected from a random superposition of linear waves \cite{nicolleau_turbulent_2000, sujovolsky_vertical_2018}. Indeed, and as already mentioned, the early time displacements are ballistic in time, have amplitudes that grow with $A$, and saturate in a time proportional to the inverse frequency of the dominant waves. However, in the horizontal plane we see three different regimes. At short times, we have a ballistic regime (again with amplitudes that grow with $A$, and which in this case are larger than what can be expected from ballistic motion generated by the mean large-scale flow). At intermediate times, we observe a saturation of the mean quadratic displacement. And later, we see a slightly slower than ballistic growth. These three regimes are separated by two characteristic times: a first time $t_{1} \approx 0.15$ s associated with the inverse of half the frequency of the most energetic waves (i.e., proportional to the time it takes for these waves to reach its maximum amplitude), and a second time $t_{2} \approx 0.35$ (for experiment A10) associated with the particle velocity correlation time, as obtained directly from the correlation of the measured velocity time series. Both times are indicated by vertical lines in the left panels of Figs.~\ref{f:dr} and \ref{f:DCM}.

\subsection{A simple random walk model}

\begin{figure}
\centering
\includegraphics[width=10cm]{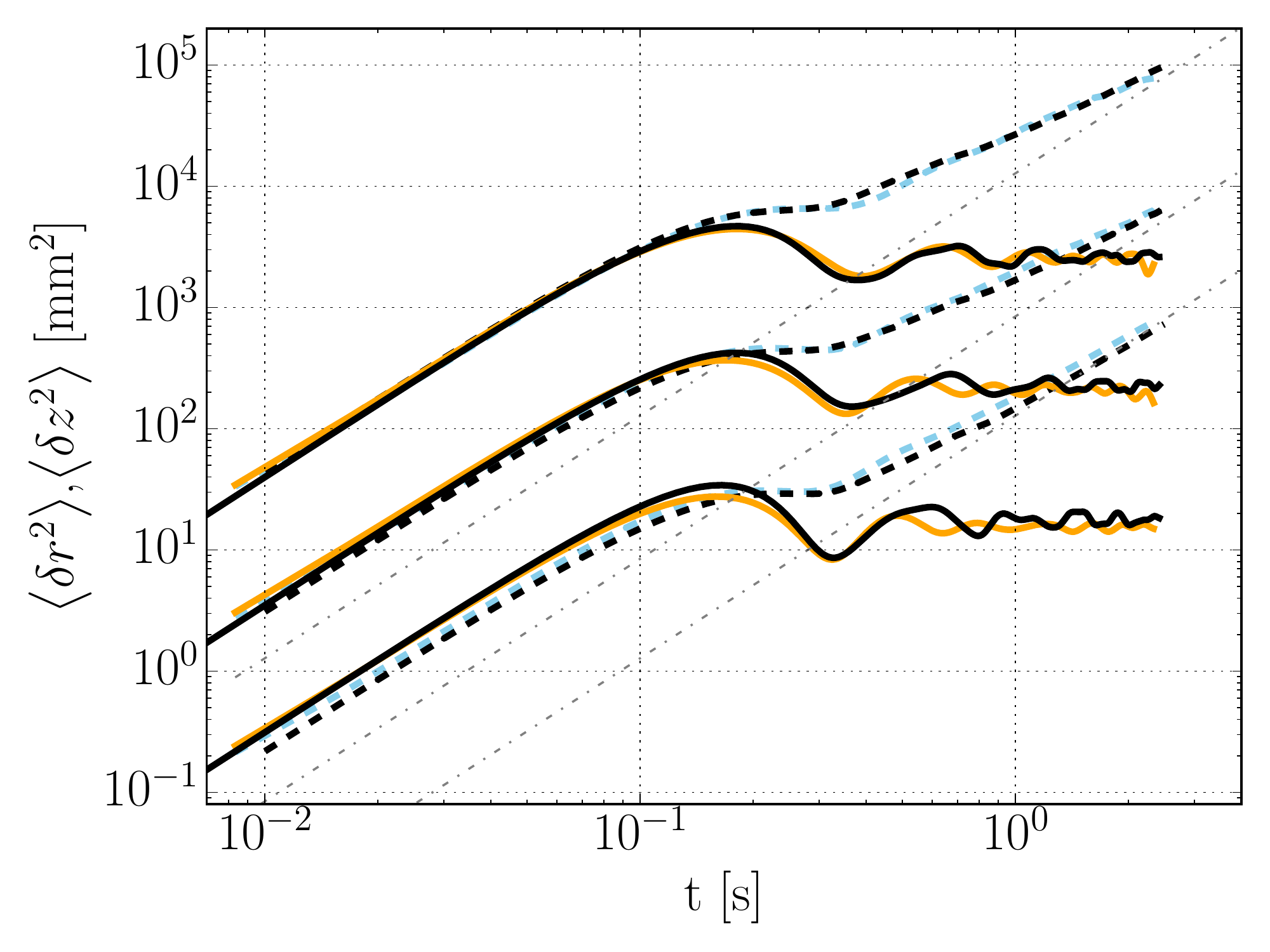}
\caption{({\it Color online}) Mean quadratic displacements in the vertical (solid lines) and horizontal directions (dashed lines), for forcing amplitudes $A = 10$, $15$, and $20$ mm (from bottom to top), as obtained from the experiments, and multiplied by an arbitrary factor to separate them vertically (color labels are as in Fig.~\ref{f:DCM}). In black lines, we show the mean quadratic displacements in the vertical (solid) and horizontal directions (dashed) obtained from the model. The case with $A=5$ is similar and not shown for simplicity. The grey dash-dotted lines indicate a ballistic displacement of the particles predicted solely from the general circulation in each of the experiments. Note that for all times these ballistic displacements are insufficient to explain the observations, and only at late times the observed displacements approach the predictions from the mean flow.}
\label{f:DCM_mod}
\end{figure}

Based on the experimental results described previously, in this section we build a model that reproduces the evolution of the particles' dispersion in both the horizontal and the vertical directions. This model will help us separate the three contributions to the displacement of floaters that were already evident from the experimental observations described above: the effect of the waves, of short-lived horizontal eddies of mid-size (i.e., smaller than the size of the vessel, although not necessarily with a single scale), and of the mean large-scale circulation of the size of the domain that also develops in the experiment. The model is based on previous models for particle dispersion developed for isotropic and homogeneous turbulence \cite{rast11, rast16}, and extended later for the coexistence of waves and eddies in stably stratified turbulence \cite{sujovolsky_single-particle_2017, sujovolsky_vertical_2018}. 

In the vertical direction we assume floaters move together with the fluid surface, whose evolution can be approximated pointwise as a random superposition of waves (i.e., of harmonic functions). Thus, the linear vertical displacement for each particle can be written as
\begin{equation}
  \delta z_i(t) = z_i(t) - z_i(0) = \sum_{\omega} A_{0} \omega^{-1} \left[
    \cos(\omega t + \phi_{\omega,i}) - \cos(\phi_{\omega,i}) \right] ,
  \label{eq:displacement}
\end{equation}
where $A_{0}$ is the amplitude of the waves, and $\phi_{\omega,i}$ is a random phase for the $i$-th particle and the frequency $\omega$ (note we can absorb inside this phase any spatial dependence associated with travelling waves). Following the observations of the vertical velocity spectrum in Fig.~\ref{f:Ef}, the sum of frequencies in Eq.~(\ref{eq:displacement}) was done over uniformly distributed random frequencies between $2$ and $4$ Hz (as the observed vertical kinetic energy spectrum has a peak around these frequencies), and with amplitudes $A_{0}=(2\left<v_{z}^{2} \right>/N_{\omega})^{1/2}$ [which results in a flat kinetic energy spectrum and in an r.m.s.~vertical velocity compatible with the experiments, where $\left<v_{z}^{2} \right>$ is the mean quadratic velocity already reported, and $N_{\omega}$ is the number of random waves considered in the sum in Eq.~(\ref{eq:displacement}); the dependence of the amplitudes in the sum as $\omega^{-1}$ results from integrating in time the constant amplitude of all harmonic modes in the velocity to get spatial displacements]. From $\delta z_i(t)$, the mean quadratic vertical displacement can finally be obtained as
\begin{equation}
\left <\delta z^2 \right>(t) = \left< \delta z_i^2 \right> (t),
\end{equation}
where the average on the r.h.s.~is done over an ensemble of particles. The results of this simple model for the vertical displacements can be seen in Fig.~\ref{f:DCM_mod}, where the model is compared against the experimental results. The random superposition of waves correctly reproduces the ballistic growth at early time, the time of saturation of $\left <\delta z^2 \right>(t)$, and the fluctuations around a mean value observed at later times.

We now proceed to consider the case of horizontal displacements. Here the waves are not sufficient to capture the dynamics in the three observed regimes. Following the observations, we consider a model with three components: (1) Drag of the floaters by the waves, (2) trapping by short-lived eddies, and (3) advection by a mean large-scale circulation. For the waves we consider a random superposition of harmonic functions as that given by Eq.~(\ref{eq:displacement}), but we now assume these waves displace the particles horizontally a distance $\delta r_i^{(\textrm{wav})}(t)$ with an associated mean velocity $v_s$ (instead of $v_z$), which is caused by the Stokes drift. To estimate the r.m.s.~value of $v_s$ we consider the Craik-Leibovich approximation \cite{craik76,craik82} (or equivalently, an averaged Lagrangian approximation \cite{andrews78,holm99}), leading to
\begin{equation}
{\bf v}_{s}=\left< \int {\bf v}_{\omega}\cdot{\boldsymbol \nabla}{\bf v}_{\omega} \, dt \right>,
\label{eq:stokes}
\end{equation}
where ${\bf v}_{\omega}$ is the velocity field associated to the gravity waves. If we estimate $v_{\omega}$ as the propagation velocity of a travelling wave, then from Eq.~(\ref{eq:stokes}), by means of dimensional analysis, and using the dispersion relation for deep water waves, we can estimate $v_s \sim k v_{z}^{2} T \sim 2\pi v_{z}^{2}/c$, where $T \approx 0.33$ s is the period of the most energetic waves (with $f=3$ Hz), $k$ is their wave number, and $c\approx 470$ mm s$^{-1}$ is their phase velocity obtained from the dispersion relation. For each experiment, the Stokes drift velocity obtained from this approximation is of the order of the observed values for $\sigma_{v_{x}}$ and $\sigma_{v_{y}}$ in table~\ref{tab:vel_accel_vs_forcing}, which is consistent with the fact that fast oscillations observed in the particles' velocities (and displacements) should be associated with drag by gravity waves.

To the displacement resulting from the waves, a continuous time random walk (CTRW) process was added to take into account trapping and displacement of the floaters by short-lived eddies, following \cite{rast16, sujovolsky_single-particle_2017}. At each step $t$, a particle is trapped and advected for a time $t_t$ by an eddy of radius $r_{t}$ with a velocity $v_{t}$. The time interval $t_{t}$ during which a particle is trapped in each step is a random variable uniformly distributed between $0$ and $0.35$ s (the observed floaters' velocity correlation time). The radius of the eddies $r_{t}$ that trap the floaters are also randomly distributed, with a distribution following Kolmogorov scaling $P(r_{t})\sim r^{4/3}$ for $r\leq L$ and $0$ for $r>L$, where $L$ is a characteristic length which we associate to the amplitude of the wavemakers' displacement $A$. Note that as a result, we consider a multiscale superposition of eddies that can trap the particles, each with a trapping (or living) time $t_{t}$; if some events of trapping can also result from superposition of long or quasi-standing waves (see, e.g., \cite{denissenko2006waves}) we will assume that our simple CTRW process can also mimic its result). Finally, the distribution of the random velocities of the eddies $v_{t}$ is given by a Rayleigh distribution with mean value $\left<v_{t}\right>=L/T$, which in practice is also close to $f_{\textrm{max}} L$ where $f_{\textrm{max}}$ is the frequency where the observed Lagrangian vertical velocity spectrum has its maximum ($\approx 3$ Hz). In each step, the floaters displace in a circle until, after the time $t_{t}$, they are trapped by another eddy. Thus, the net horizontal displacement for each particle, resulting from the trapping by the eddies in each step, is $\delta r^{\textrm{CTRW}}_i = 2r_t |\sin(\theta_t)|$, where $\theta_t = v_{t} t_t/r_t$ is the central angle of motion of the particle while trapped.

\begin{figure}
\centering
\includegraphics[width=10cm]{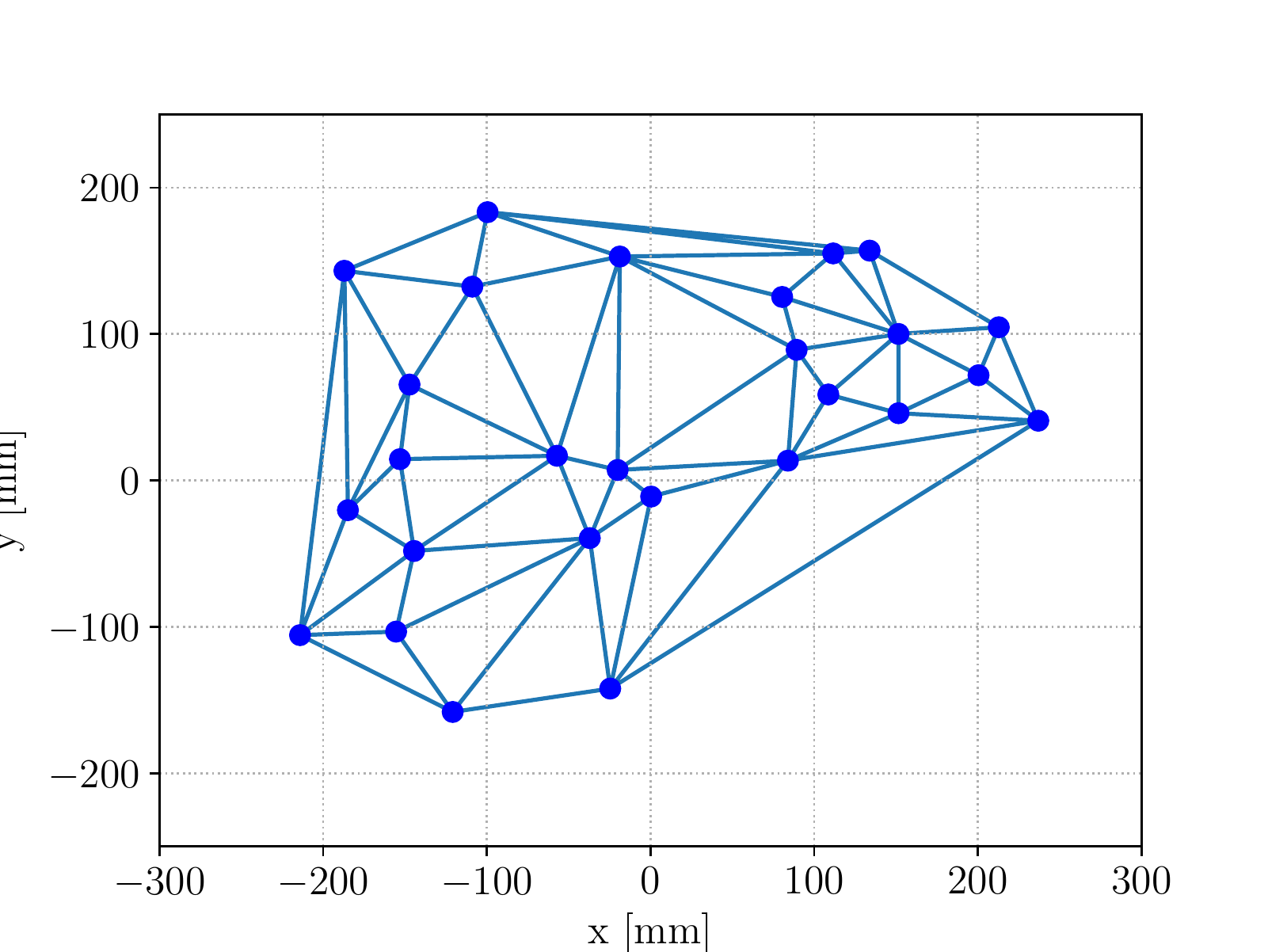}
\caption{({\it Color online}) Delaunay tessellation for an instantaneous configuration of floaters, using a small number of particles in experiment A10.}
\label{f:cluster}
\end{figure}

Finally, to model the contribution from the large-scale coherent circulation in the vessel, we model that circulation using a velocity
\begin{equation}
\centering
{\bf v}^{\textrm{(circ)}}(x,y) = u_{0} \sin \left( \frac{2\pi x}{L_{x}} \right) \cos \left( \frac{2\pi y}{L_{y}} \right) {\hat x} - u_{0} \frac{L_{y}}{L_{x}}  \cos \left( \frac{2\pi x}{L_{x}} \right) \sin \left( \frac{2\pi y}{L_{y}} \right) {\hat y},
\end{equation}
where $L_{x}$ and $L_{y}$ are the lengths of the tank in each direction and $u_{0}$ is the characteristic velocity of the circulation obtained from simple inspection of the particle trajectories (see Fig.~\ref{f:xyz}; in practice, $u_0$ is in all cases one order of magnitude smaller than the observed values of $v_x$, $v_y$, $\sigma_{v_{x}}$, and $\sigma_{v_{y}}$ in table~\ref{tab:vel_accel_vs_forcing}). The advection of randomly placed particles by this circulation $\delta r_i^{\textrm{(circ)}}$ is obtained from direct integration.

The total mean quadratic horizontal displacement is finally given by
\begin{equation}
\left< \delta r^2 \right>(t) = \left< \left( \delta r_i^{\textrm{(wav)}} +\delta r_i^{\textrm{(CTRW)}} + \delta r_i^{\textrm{(circ)}} \right)^2 \right> ,
\end{equation}
where again the l.h.s.~average is computed over an ensemble of particles.

Figure \ref{f:DCM_mod} shows the mean squared horizontal displacements obtained from this model together with the results from the experiments, for the cases with $A=10$, $15$, and $20$ mm (the case with $A=5$ is similar and not shown for simplicity). Note that while the curves are multiplied by arbitrary factors to separate them vertically and help the comparison, the same factors were used for the model and the experimental data in each case, and thus there is no prefactor adjusted to get the correct amplitudes. The model is able to capture the three regimes, and further confirms that these three physical effects are responsible for the general evolution of the particle displacements. Indeed, each component of the model (drag by the waves, trapping by short-lived eddies, and advection by the large-scale mean circulation) is needed to reproduce the three regimes observed in the experiments: The effect of the waves is required to explain the early ballistic regime, trapping by short-time eddies captures the mid-time saturation, and the late time dispersion cannot be reproduced without taking into account the slow advection by the mean circulation (whose contribution is, however, negligible at early times as also shown in Fig. \ref{f:DCM_mod}). Note that the model presented here is not intended as a fundamental model for the transport of floaters in free surface flows (some choices, although based on the experimental observations, are somewhat arbitrary), but as a proof of concept pointing at the basic physical ingredients at work that result in the dispersion seen in Figs.~\ref{f:dr} and \ref{f:DCM}.

\section{Preferential concentration of floaters \label{sec:clus}}

\begin{figure}
\centering
\includegraphics[width=8.5cm]{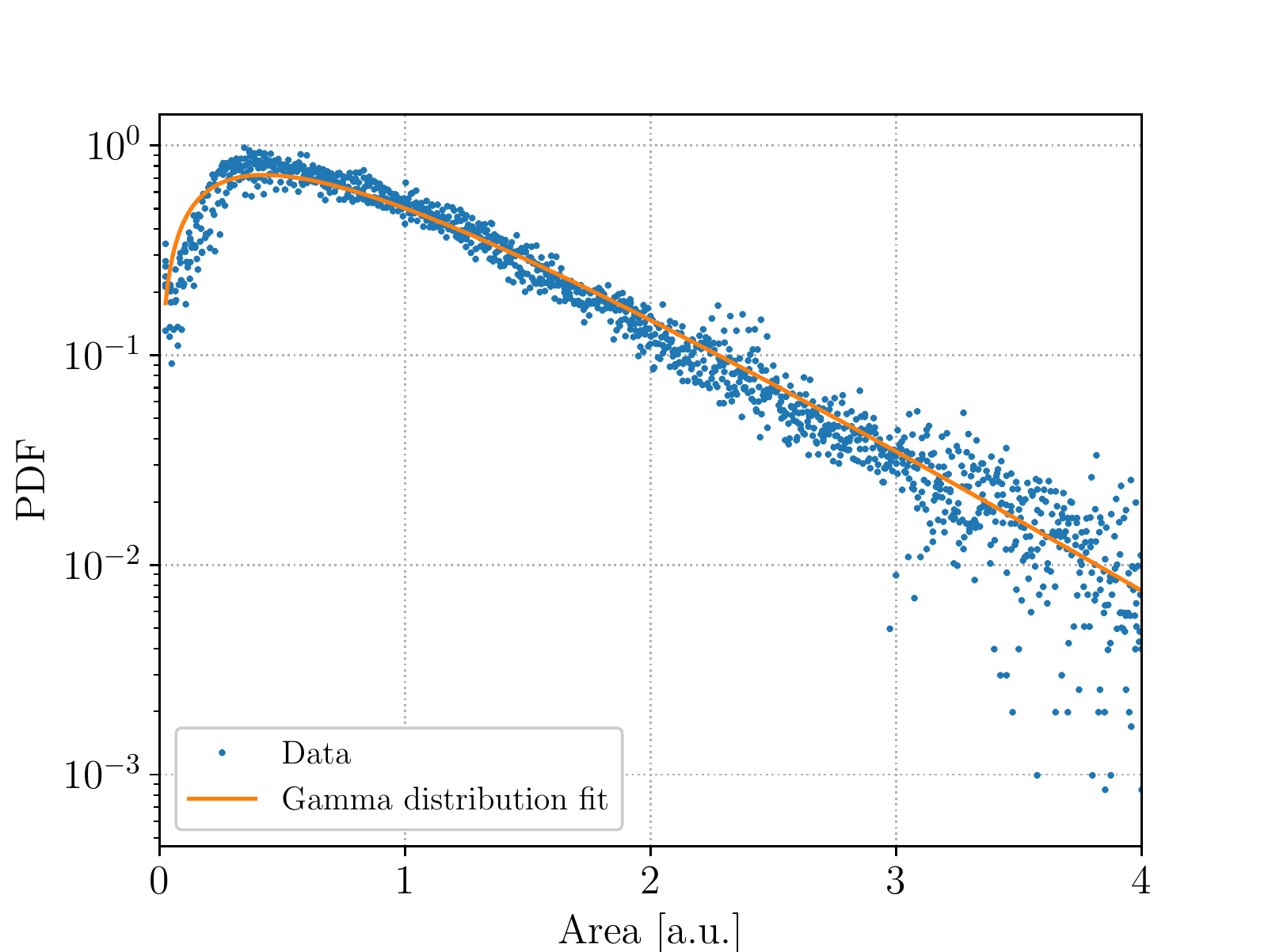}
\includegraphics[width=8.5cm]{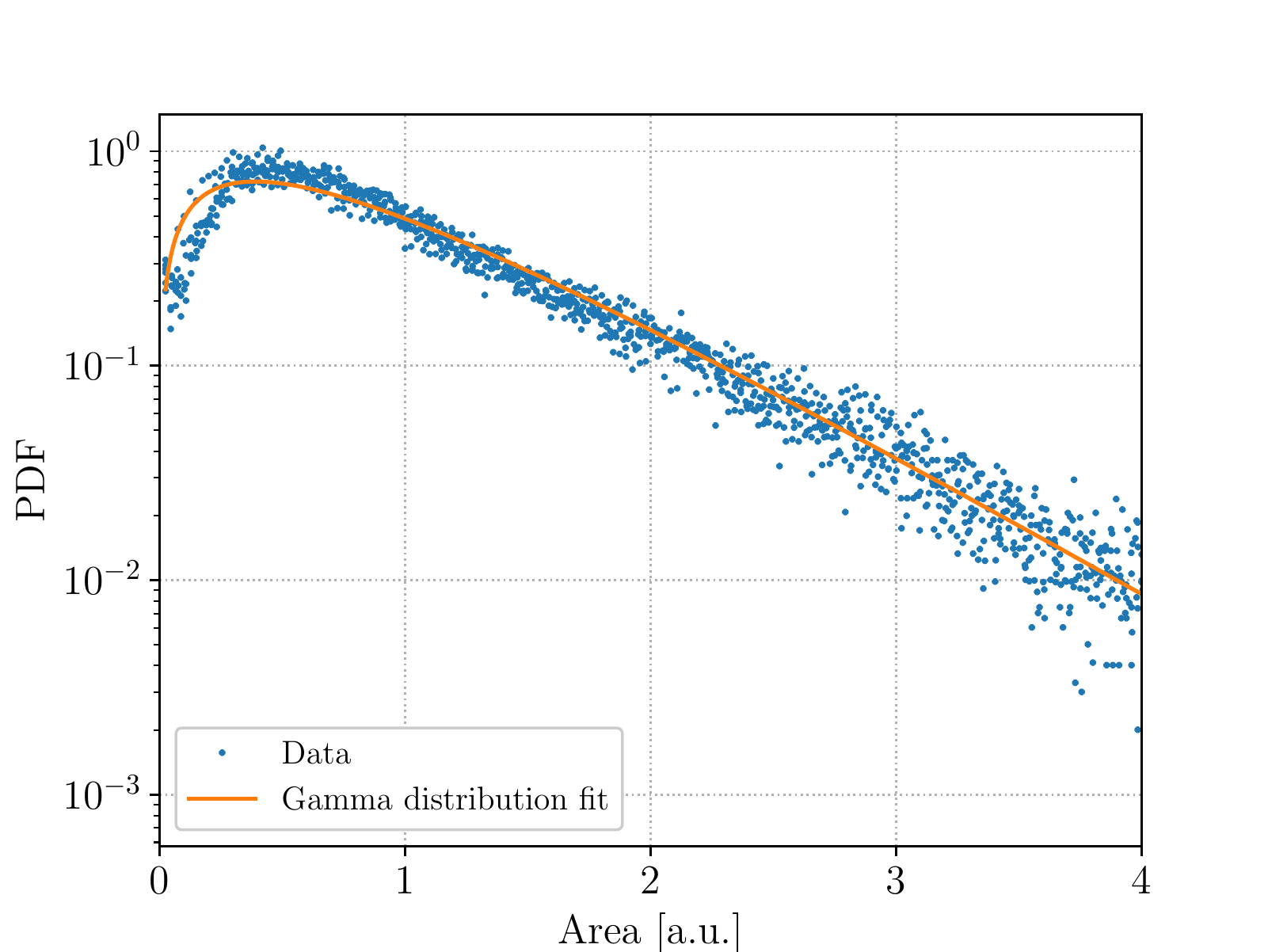}
\caption{({\it Color online}) PDFs of the areas of the triangular cells resulting from the Delaunay tessellation, in experiment A10 ({\it left}) and A20 ({\it right}). As a reference, a best fit using a Gamma distribution function is shown in both panels.}
\label{f:area}
\end{figure}

Finally, as a first approach to the study of multi-particle statistics of floaters dispersed by free surface flows, we briefly study the phenomenon of preferential concentration and of formation of clusters in the experiments: Does this flow exhibit preferential concentration of the floaters, or are they randomly distributed in space? To answer this question we consider the position of particles in the $x$-$y$ plane and perform a Delaunay tessellation on the images frame by frame, using the position of the particles in the plane as vertices. The Delaunay tessellation (or, other tessellations such as the Vorono\"i tessellation \cite{obligado14}, which in fact is the dual of the Delaunay tessellation) allow quantification of the mean area occupied by the particles, and can be used instead of measurements of mean particle density to characterize inhomogeneities in the particles distribution (see, e.g., \cite{denissenko2006waves}). An example of the Delaunay tessellation for one frame in experiment A10, and for just a few particles, can be seen in Fig.~\ref{f:cluster}.

In order to study cluster formation we computed the PDFs of the areas of the triangles, normalized by the mean of the distribution, as shown in Fig.~\ref{f:area} for experiments A10 and A20. Interestingly, the shape of the PDFs seems insensitive to the strength of the forcing $A$. Moreover, it can be shown \cite{aste08} that for a random distribution of points, the distribution of areas $S$ (properly made dimensionless) obtained from a Delaunay tessellation follows the Gamma distribution,
\begin{equation}
P(S) = \frac{b^a}{\Gamma(a)} S^{b-1} e^{-bS},
\end{equation}
where $\Gamma$ is the gamma function, $a=\left<S \right>^2/\sigma_S^2$, $a=\left<S \right>/\sigma_S^2$, and $\sigma_S$ is the standard deviation of $S$. For uniformly distributed random points, $a=1$ and $b=1/\left<S \right>$ \cite{gutierrez_clustering_2016}. Figure \ref{f:area} shows a best fit to the observed PDFs using the Gamma distribution. The PDFs are compatible with this distribution with $a = (1.7 \pm 0.3)$ for all forcing amplitudes, a signature of a relative excess of larger areas, and thus indicating the presence of clusters in the flow. Using the criteria proposed in \cite{gutierrez_clustering_2016} for data that follows a Gamma distribution, we further determined a critical dimensionless area $A_{c} = 0.23$. Thus, only floaters forming a triangle of dimensionless area smaller than $A_c$ (i.e., to the left of the peak of the distribution) belong to a cluster.

Based on the PDFs we can conclude that surface wave turbulence, albeit with possible contributions from eddies and the large-scale circulation in the vessel, generates clustering of floaters. However, the reasons behind the observed preferential concentration are not easy to disentangle. The independence of the observed PDFs with the forcing amplitude $A$ (which controls the strength of the large-scale mean circulation and of the horizontal eddies, and affects the mean quadratic particle displacements as discussed in previous sections), suggests that mid-size eddies and the large-scale circulation do not play a dominant role in the observed clustering. Particles in the surface of the fluid can also experience effective compressibility effects, which can result in floaters being attracted to contracting regions of the flow \cite{boffetta04}, albeit in practice a weak effective compressibility reduces the preferential concentration. Finally, capillarity can also result in the agglomeration of small particles (as capillarity makes particles with similar wetting attractive) \cite{dalbe11}; however in our case the particles' diameter is of the order of the gravity-capillary crossover length $\lambda_C$, and we do not observe particles sticking together in the experiments). We are thus left with the effect of the Stokes drift by traveling surface waves, of standing waves, and of the random superposition of waves expected in a wave turbulent flow. It has been shown that inhomogeneities of a transported quantity can grow exponentially in wave turbulence \cite{balk2004growth}, although for a potential flow the growth is slow (note however that our flow has solenoidal contributions). More probably, a faster mechanism for the accumulation is the break down of Archimedes' law associated with capillarity, which makes the floaters inertial \cite{falkovich05, denissenko2006waves}. Such preferential concentration was observed in experiments of wave turbulence with small particles in \cite{denissenko2006waves}, where the accumulation was studied using particle concentrations.

\section{Conclusions \label{sec:conc}}

While in recent years there have been significant advances in the experimental study of surface wave turbulence in the Eulerian framework \cite{cobelli11, berhanu13, aubourg15, paquier15, clark15, deike2015role, paquier16}, the study of this problem in the laboratory from the point of view of Lagrangian trajectories or of buoyant inertial tracers has received significantly less attention \cite{francois2014three, punzmann2014generation}. Here we presented laboratory experiments of surface wave turbulence excited by paddles in the deep water regime, with the surface of the fluid seeded with buoyant particles or floaters. For different forcing amplitudes, all resulting in r.m.s.~wave steepness (which measures the strength of non-linearities) smaller than 3\%, we performed an ensemble of 7 experiments for each forcing, and simultaneously measured particle trajectories for 23 s and for an average of 52 floaters in each experiment realization. The displacement, velocity, and acceleration of each floater was obtained using particle tracking velocimetry with two ultra-fast cameras and a 120 Hz acquisition frequency. We studied the statistics of the floaters' velocity and acceleration, the velocity power spectrum, the mean squared vertical and horizontal displacements, and the formation of clusters of particles using a Delaunay tessellation.

The statistics of all components of the floaters' velocity and acceleration is close to Gaussian, with no dependence on the forcing amplitude, at least for the range of values considered in our experiments. The observed kinetic energy frequency spectra are reminiscent of those found in oceanic measurements using buoys \cite{garrett_space-time_1975, dasaro_lagrangian_2000}, and display a rather shallow or flat spectrum for frequencies smaller than those of the surface waves excited by the paddles ($< 3$ Hz), a peak around this frequency, and a power law decay compatible with $E(f) \sim f^{-4.5}$ for higher frequencies. While for low frequencies ($\lesssim 1$ Hz) the power spectrum of the vertical velocity $v_z$ is flat, a shallow dependence with the frequency $f$ is observed in the power spectrum of the horizontal velocity components $v_x$ and $v_y$. Given the length of our vessel of $198$ cm, for frequencies $f \lesssim 10^{-1}$ s$^{-1}$ no waves can be expected to be associated with these excitations, and thus such low frequency excitations observed in the power spectrum of $v_x$ and $v_y$ can only be associated with the contribution from horizontal eddies and a slow advection by a large-scale circulation that builds over time in the entire vessel. The squared amplitude of these modes is, however, from 2 to 20 times smaller (depending on the frequency and on the experiment) than the peak squared amplitude of the forced wave modes with $f\approx 3$ Hz.

The measured mean squared vertical and horizontal displacements of the particles display two or three separate behaviors at early, intermediate, and late times. On the one hand, the mean squared vertical displacement shows a ballistic growth at early times (before the half period of the strongest waves), followed by a saturation at a level proportional to the mean squared amplitude of the waves. On the other hand, the mean squared horizontal displacement shows a ballistic growth at early times, a saturation at intermediate times, and a second growth with time at late times (after the floaters' velocity correlation time is reached). We also presented a simple model for these observed displacements. For vertical displacements, assuming that the vertical position of the particles follows a random superposition of harmonic waves suffices to quantitatively reproduce the observations. For horizontal displacements the model is built upon: (1) drag by a random superposition of waves, (2) a continuous-time random walk process associated to eddy turbulence, and (3) a slow deterministic displacement. This model properly captures the three behaviors seen in the experimental data, and successfully associates each behavior with a physical process: the ballistic growth of the dispersion at early times is dominated by the effect of the waves, the saturation at intermediate times is caused by the trapping of the floaters by short-lived mid-size eddies, and the growth at late times results from the advection of the floaters by the large-scale mean circulation in the vessel. The success of this simple model in capturing the dispersion at different times not only allowed us to disentangle the different physical effects contributing to the dispersion of particles in the flow, but also confirms the utility of continuous time random walk models \cite{rast11, rast16, sujovolsky_single-particle_2017, sujovolsky_vertical_2018} to generally describe the statistical displacement of particles in turbulent flows. However, it is important to note that for these experiments the models were mostly used as a way to identify dominant physical processes acting on different time scales, as parameters were obtained or adjusted from the experimental observations.

Finally, and as a first approach at studying multi-particle statistics in these flows, the Delaunay tessellation allowed us to identify the formation of clusters. Independently of the amplitude of the forcing, and with criteria used in previous studies of preferential concentration of particles (see, e.g., \cite{gutierrez_clustering_2016}), the statistical results indicate that clusters of floaters are indeed generated in the experiments. Overall the results show that valuable information on free surface flows, as well as on their transport and dispersion properties, can be obtained from measurements from particle tracking velocimetry using buoyant particles, richly supplementing information accessed in previous studies using Eulerian or surface deformation measurements.

\begin{acknowledgments}
The authors acknowledge finantial support from grants UBACYT No.~20020170100508BA and PICT No.~2015-3530.

\end{acknowledgments}
\bibliography{ms}

\begin{thebibliography}{52}%
\makeatletter
\providecommand \@ifxundefined [1]{%
 \@ifx{#1\undefined}
}%
\providecommand \@ifnum [1]{%
 \ifnum #1\expandafter \@firstoftwo
 \else \expandafter \@secondoftwo
 \fi
}%
\providecommand \@ifx [1]{%
 \ifx #1\expandafter \@firstoftwo
 \else \expandafter \@secondoftwo
 \fi
}%
\providecommand \natexlab [1]{#1}%
\providecommand \enquote  [1]{``#1''}%
\providecommand \bibnamefont  [1]{#1}%
\providecommand \bibfnamefont [1]{#1}%
\providecommand \citenamefont [1]{#1}%
\providecommand \href@noop [0]{\@secondoftwo}%
\providecommand \href [0]{\begingroup \@sanitize@url \@href}%
\providecommand \@href[1]{\@@startlink{#1}\@@href}%
\providecommand \@@href[1]{\endgroup#1\@@endlink}%
\providecommand \@sanitize@url [0]{\catcode `\\12\catcode `\$12\catcode
  `\&12\catcode `\#12\catcode `\^12\catcode `\_12\catcode `\%12\relax}%
\providecommand \@@startlink[1]{}%
\providecommand \@@endlink[0]{}%
\providecommand \url  [0]{\begingroup\@sanitize@url \@url }%
\providecommand \@url [1]{\endgroup\@href {#1}{\urlprefix }}%
\providecommand \urlprefix  [0]{URL }%
\providecommand \Eprint [0]{\href }%
\providecommand \doibase [0]{http://dx.doi.org/}%
\providecommand \selectlanguage [0]{\@gobble}%
\providecommand \bibinfo  [0]{\@secondoftwo}%
\providecommand \bibfield  [0]{\@secondoftwo}%
\providecommand \translation [1]{[#1]}%
\providecommand \BibitemOpen [0]{}%
\providecommand \bibitemStop [0]{}%
\providecommand \bibitemNoStop [0]{.\EOS\space}%
\providecommand \EOS [0]{\spacefactor3000\relax}%
\providecommand \BibitemShut  [1]{\csname bibitem#1\endcsname}%
\let\auto@bib@innerbib\@empty
\bibitem [{\citenamefont {Zakharov}\ \emph {et~al.}(2012)\citenamefont
  {Zakharov}, \citenamefont {L'vov},\ and\ \citenamefont
  {Falkovich}}]{zakharov12}%
  \BibitemOpen
  \bibfield  {author} {\bibinfo {author} {\bibfnamefont {V.~E.}\ \bibnamefont
  {Zakharov}}, \bibinfo {author} {\bibfnamefont {V.~S.}\ \bibnamefont {L'vov}},
  \ and\ \bibinfo {author} {\bibfnamefont {G.}~\bibnamefont {Falkovich}},\
  }\href@noop {} {\emph {\bibinfo {title} {Kolmogorov spectra of turbulence
  {I}: {W}ave turbulence}}}\ (\bibinfo  {publisher} {Springer Science \&
  Business Media},\ \bibinfo {year} {2012})\BibitemShut {NoStop}%
\bibitem [{\citenamefont {Newell}\ and\ \citenamefont
  {Rumpf}(2011)}]{newell11}%
  \BibitemOpen
  \bibfield  {author} {\bibinfo {author} {\bibfnamefont {A.~C.}\ \bibnamefont
  {Newell}}\ and\ \bibinfo {author} {\bibfnamefont {B.}~\bibnamefont {Rumpf}},\
  }\bibfield  {title} {\enquote {\bibinfo {title} {Wave turbulence},}\
  }\href@noop {} {\bibfield  {journal} {\bibinfo  {journal} {Annu. Rev. Fluid
  Mech.}\ }\textbf {\bibinfo {volume} {43}},\ \bibinfo {pages} {59--78}
  (\bibinfo {year} {2011})}\BibitemShut {NoStop}%
\bibitem [{\citenamefont {Nazarenko}(2011)}]{nazarenko11}%
  \BibitemOpen
  \bibfield  {author} {\bibinfo {author} {\bibfnamefont {S.}~\bibnamefont
  {Nazarenko}},\ }\href@noop {} {\emph {\bibinfo {title} {Wave Turbulence}}}\
  (\bibinfo  {publisher} {Springer-Verlag},\ \bibinfo {year}
  {2011})\BibitemShut {NoStop}%
\bibitem [{\citenamefont {Cobelli}\ \emph {et~al.}(2009)\citenamefont
  {Cobelli}, \citenamefont {Petitjeans}, \citenamefont {Maurel}, \citenamefont
  {Pagneux},\ and\ \citenamefont {Mordant}}]{cobelli09}%
  \BibitemOpen
  \bibfield  {author} {\bibinfo {author} {\bibfnamefont {P.}~\bibnamefont
  {Cobelli}}, \bibinfo {author} {\bibfnamefont {P.}~\bibnamefont {Petitjeans}},
  \bibinfo {author} {\bibfnamefont {A.}~\bibnamefont {Maurel}}, \bibinfo
  {author} {\bibfnamefont {V.}~\bibnamefont {Pagneux}}, \ and\ \bibinfo
  {author} {\bibfnamefont {N.}~\bibnamefont {Mordant}},\ }\bibfield  {title}
  {\enquote {\bibinfo {title} {Space-time resolved wave turbulence in a
  vibrating plate},}\ }\href@noop {} {\bibfield  {journal} {\bibinfo  {journal}
  {Phys. Rev. Lett.}\ }\textbf {\bibinfo {volume} {103}},\ \bibinfo {pages}
  {204301} (\bibinfo {year} {2009})}\BibitemShut {NoStop}%
\bibitem [{\citenamefont {Mordant}(2010)}]{mordant10}%
  \BibitemOpen
  \bibfield  {author} {\bibinfo {author} {\bibfnamefont {N.}~\bibnamefont
  {Mordant}},\ }\bibfield  {title} {\enquote {\bibinfo {title} {Fourier
  analysis of wave turbulence in a thin elastic plate},}\ }\href@noop {}
  {\bibfield  {journal} {\bibinfo  {journal} {Eur. Phys. J. B}\ }\textbf
  {\bibinfo {volume} {76}},\ \bibinfo {pages} {537--545} (\bibinfo {year}
  {2010})}\BibitemShut {NoStop}%
\bibitem [{\citenamefont {Cobelli}\ \emph {et~al.}(2011)\citenamefont
  {Cobelli}, \citenamefont {Przadka}, \citenamefont {Petitjeans}, \citenamefont
  {Lagubeau}, \citenamefont {Pagneux},\ and\ \citenamefont
  {Maurel}}]{cobelli11}%
  \BibitemOpen
  \bibfield  {author} {\bibinfo {author} {\bibfnamefont {P.}~\bibnamefont
  {Cobelli}}, \bibinfo {author} {\bibfnamefont {A.}~\bibnamefont {Przadka}},
  \bibinfo {author} {\bibfnamefont {P.}~\bibnamefont {Petitjeans}}, \bibinfo
  {author} {\bibfnamefont {G.}~\bibnamefont {Lagubeau}}, \bibinfo {author}
  {\bibfnamefont {V.}~\bibnamefont {Pagneux}}, \ and\ \bibinfo {author}
  {\bibfnamefont {A.}~\bibnamefont {Maurel}},\ }\bibfield  {title} {\enquote
  {\bibinfo {title} {Different regimes for water wave turbulence},}\
  }\href@noop {} {\bibfield  {journal} {\bibinfo  {journal} {Phys. Rev. Lett.}\
  }\textbf {\bibinfo {volume} {107}},\ \bibinfo {pages} {214503} (\bibinfo
  {year} {2011})}\BibitemShut {NoStop}%
\bibitem [{\citenamefont {Berhanu}\ and\ \citenamefont
  {Falcon}(2013)}]{berhanu13}%
  \BibitemOpen
  \bibfield  {author} {\bibinfo {author} {\bibfnamefont {M.}~\bibnamefont
  {Berhanu}}\ and\ \bibinfo {author} {\bibfnamefont {E.}~\bibnamefont
  {Falcon}},\ }\bibfield  {title} {\enquote {\bibinfo {title}
  {Space-time-resolved capillary wave turbulence},}\ }\href@noop {} {\bibfield
  {journal} {\bibinfo  {journal} {Phys. Rev. E}\ }\textbf {\bibinfo {volume}
  {87}},\ \bibinfo {pages} {033003} (\bibinfo {year} {2013})}\BibitemShut
  {NoStop}%
\bibitem [{\citenamefont {Aubourg}\ and\ \citenamefont
  {Mordant}(2015)}]{aubourg15}%
  \BibitemOpen
  \bibfield  {author} {\bibinfo {author} {\bibfnamefont {Q.}~\bibnamefont
  {Aubourg}}\ and\ \bibinfo {author} {\bibfnamefont {N.}~\bibnamefont
  {Mordant}},\ }\bibfield  {title} {\enquote {\bibinfo {title} {Nonlocal
  resonances in weak turbulence of gravity-capillary waves},}\ }\href@noop {}
  {\bibfield  {journal} {\bibinfo  {journal} {Phys. Rev. Lett.}\ }\textbf
  {\bibinfo {volume} {114}},\ \bibinfo {pages} {144501} (\bibinfo {year}
  {2015})}\BibitemShut {NoStop}%
\bibitem [{\citenamefont {Clark~di Leoni}\ \emph {et~al.}(2015)\citenamefont
  {Clark~di Leoni}, \citenamefont {Cobelli},\ and\ \citenamefont
  {Mininni}}]{clark15}%
  \BibitemOpen
  \bibfield  {author} {\bibinfo {author} {\bibfnamefont {P.}~\bibnamefont
  {Clark~di Leoni}}, \bibinfo {author} {\bibfnamefont {P.~J.}\ \bibnamefont
  {Cobelli}}, \ and\ \bibinfo {author} {\bibfnamefont {P.~D.}\ \bibnamefont
  {Mininni}},\ }\bibfield  {title} {\enquote {\bibinfo {title} {The
  spatio-temporal spectrum of turbulent flows},}\ }\href@noop {} {\bibfield
  {journal} {\bibinfo  {journal} {Eur. Phys. J. E}\ }\textbf {\bibinfo {volume}
  {38}},\ \bibinfo {pages} {136} (\bibinfo {year} {2015})}\BibitemShut
  {NoStop}%
\bibitem [{\citenamefont {Deike}\ \emph {et~al.}(2015)\citenamefont {Deike},
  \citenamefont {Miquel}, \citenamefont {Guti{\'e}rrez}, \citenamefont {Jamin},
  \citenamefont {Semin}, \citenamefont {Berhanu}, \citenamefont {Falcon},\ and\
  \citenamefont {Bonnefoy}}]{deike2015role}%
  \BibitemOpen
  \bibfield  {author} {\bibinfo {author} {\bibfnamefont {L.}~\bibnamefont
  {Deike}}, \bibinfo {author} {\bibfnamefont {B.}~\bibnamefont {Miquel}},
  \bibinfo {author} {\bibfnamefont {P.}~\bibnamefont {Guti{\'e}rrez}}, \bibinfo
  {author} {\bibfnamefont {T.}~\bibnamefont {Jamin}}, \bibinfo {author}
  {\bibfnamefont {B.}~\bibnamefont {Semin}}, \bibinfo {author} {\bibfnamefont
  {M.}~\bibnamefont {Berhanu}}, \bibinfo {author} {\bibfnamefont
  {E.}~\bibnamefont {Falcon}}, \ and\ \bibinfo {author} {\bibfnamefont
  {F.}~\bibnamefont {Bonnefoy}},\ }\bibfield  {title} {\enquote {\bibinfo
  {title} {Role of the basin boundary conditions in gravity wave turbulence},}\
  }\href@noop {} {\bibfield  {journal} {\bibinfo  {journal} {J. Fluid Mech.}\
  }\textbf {\bibinfo {volume} {781}},\ \bibinfo {pages} {196--225} (\bibinfo
  {year} {2015})}\BibitemShut {NoStop}%
\bibitem [{\citenamefont {Paquier}\ \emph {et~al.}(2015)\citenamefont
  {Paquier}, \citenamefont {Moisy},\ and\ \citenamefont {Rabaud}}]{paquier15}%
  \BibitemOpen
  \bibfield  {author} {\bibinfo {author} {\bibfnamefont {A.}~\bibnamefont
  {Paquier}}, \bibinfo {author} {\bibfnamefont {F.}~\bibnamefont {Moisy}}, \
  and\ \bibinfo {author} {\bibfnamefont {M.}~\bibnamefont {Rabaud}},\
  }\bibfield  {title} {\enquote {\bibinfo {title} {Surface deformations and
  wave generation by wind blowing over a viscous liquid},}\ }\href@noop {}
  {\bibfield  {journal} {\bibinfo  {journal} {Phys. Fluids}\ }\textbf {\bibinfo
  {volume} {27}},\ \bibinfo {pages} {122103} (\bibinfo {year}
  {2015})}\BibitemShut {NoStop}%
\bibitem [{\citenamefont {Paquier}\ \emph {et~al.}(2016)\citenamefont
  {Paquier}, \citenamefont {Moisy},\ and\ \citenamefont {Rabaud}}]{paquier16}%
  \BibitemOpen
  \bibfield  {author} {\bibinfo {author} {\bibfnamefont {A.}~\bibnamefont
  {Paquier}}, \bibinfo {author} {\bibfnamefont {F.}~\bibnamefont {Moisy}}, \
  and\ \bibinfo {author} {\bibfnamefont {M.}~\bibnamefont {Rabaud}},\
  }\bibfield  {title} {\enquote {\bibinfo {title} {Viscosity effects in wind
  wave generation},}\ }\href@noop {} {\bibfield  {journal} {\bibinfo  {journal}
  {Phys. Rev. Fluids}\ }\textbf {\bibinfo {volume} {1}},\ \bibinfo {pages}
  {083901} (\bibinfo {year} {2016})}\BibitemShut {NoStop}%
\bibitem [{\citenamefont {Mordant}\ \emph {et~al.}(2001)\citenamefont
  {Mordant}, \citenamefont {Metz}, \citenamefont {Michel},\ and\ \citenamefont
  {Pinton}}]{mordant01}%
  \BibitemOpen
  \bibfield  {author} {\bibinfo {author} {\bibfnamefont {N.}~\bibnamefont
  {Mordant}}, \bibinfo {author} {\bibfnamefont {P.}~\bibnamefont {Metz}},
  \bibinfo {author} {\bibfnamefont {O.}~\bibnamefont {Michel}}, \ and\ \bibinfo
  {author} {\bibfnamefont {J.-F.}\ \bibnamefont {Pinton}},\ }\bibfield  {title}
  {\enquote {\bibinfo {title} {Measurement of {L}agrangian velocity in fully
  developed turbulence},}\ }\href@noop {} {\bibfield  {journal} {\bibinfo
  {journal} {Phys. Rev. Lett.}\ }\textbf {\bibinfo {volume} {87}},\ \bibinfo
  {pages} {214501} (\bibinfo {year} {2001})}\BibitemShut {NoStop}%
\bibitem [{\citenamefont {Rast}\ and\ \citenamefont {Pinton}(2011)}]{rast11}%
  \BibitemOpen
  \bibfield  {author} {\bibinfo {author} {\bibfnamefont {M.~P.}\ \bibnamefont
  {Rast}}\ and\ \bibinfo {author} {\bibfnamefont {J.-F.}\ \bibnamefont
  {Pinton}},\ }\bibfield  {title} {\enquote {\bibinfo {title} {Pair dispersion
  in turbulence: {T}he subdominant role of scaling},}\ }\href@noop {}
  {\bibfield  {journal} {\bibinfo  {journal} {Phys. Rev. Lett.}\ }\textbf
  {\bibinfo {volume} {107}},\ \bibinfo {pages} {214501} (\bibinfo {year}
  {2011})}\BibitemShut {NoStop}%
\bibitem [{\citenamefont {Rast}\ \emph {et~al.}(2016)\citenamefont {Rast},
  \citenamefont {Pinton},\ and\ \citenamefont {Mininni}}]{rast16}%
  \BibitemOpen
  \bibfield  {author} {\bibinfo {author} {\bibfnamefont {M.~P.}\ \bibnamefont
  {Rast}}, \bibinfo {author} {\bibfnamefont {J.-F.}\ \bibnamefont {Pinton}}, \
  and\ \bibinfo {author} {\bibfnamefont {P.~D.}\ \bibnamefont {Mininni}},\
  }\bibfield  {title} {\enquote {\bibinfo {title} {Turbulent transport with
  intermittency: {E}xpectation of a scalar concentration},}\ }\href@noop {}
  {\bibfield  {journal} {\bibinfo  {journal} {Phys. Rev. E}\ }\textbf {\bibinfo
  {volume} {93}},\ \bibinfo {pages} {043120} (\bibinfo {year}
  {2016})}\BibitemShut {NoStop}%
\bibitem [{\citenamefont {Salazar}\ \emph {et~al.}(2008)\citenamefont
  {Salazar}, \citenamefont {De~Jong}, \citenamefont {Cao}, \citenamefont
  {Woodward}, \citenamefont {Meng},\ and\ \citenamefont {Collins}}]{salazar08}%
  \BibitemOpen
  \bibfield  {author} {\bibinfo {author} {\bibfnamefont {J.~P. L.~C.}\
  \bibnamefont {Salazar}}, \bibinfo {author} {\bibfnamefont {J.}~\bibnamefont
  {De~Jong}}, \bibinfo {author} {\bibfnamefont {L.}~\bibnamefont {Cao}},
  \bibinfo {author} {\bibfnamefont {S.~H.}\ \bibnamefont {Woodward}}, \bibinfo
  {author} {\bibfnamefont {H.}~\bibnamefont {Meng}}, \ and\ \bibinfo {author}
  {\bibfnamefont {L.~R.}\ \bibnamefont {Collins}},\ }\bibfield  {title}
  {\enquote {\bibinfo {title} {Experimental and numerical investigation of
  inertial particle clustering in isotropic turbulence},}\ }\href@noop {}
  {\bibfield  {journal} {\bibinfo  {journal} {J. Fluid. Mech.}\ }\textbf
  {\bibinfo {volume} {600}},\ \bibinfo {pages} {245--256} (\bibinfo {year}
  {2008})}\BibitemShut {NoStop}%
\bibitem [{\citenamefont {Goto}\ and\ \citenamefont
  {Vassilicos}(2008)}]{goto08}%
  \BibitemOpen
  \bibfield  {author} {\bibinfo {author} {\bibfnamefont {S.}~\bibnamefont
  {Goto}}\ and\ \bibinfo {author} {\bibfnamefont {J.~C.}\ \bibnamefont
  {Vassilicos}},\ }\bibfield  {title} {\enquote {\bibinfo {title} {Sweep-stick
  mechanism of heavy particle clustering in fluid turbulence},}\ }\href@noop {}
  {\bibfield  {journal} {\bibinfo  {journal} {Phys. Rev. Lett.}\ }\textbf
  {\bibinfo {volume} {100}},\ \bibinfo {pages} {054503} (\bibinfo {year}
  {2008})}\BibitemShut {NoStop}%
\bibitem [{\citenamefont {Coleman}\ and\ \citenamefont
  {Vassilicos}(2009)}]{coleman09}%
  \BibitemOpen
  \bibfield  {author} {\bibinfo {author} {\bibfnamefont {S.~W.}\ \bibnamefont
  {Coleman}}\ and\ \bibinfo {author} {\bibfnamefont {J.~C.}\ \bibnamefont
  {Vassilicos}},\ }\bibfield  {title} {\enquote {\bibinfo {title} {A unified
  sweep-stick mechanism to explain particle clustering in two-and
  three-dimensional homogeneous, isotropic turbulence},}\ }\href@noop {}
  {\bibfield  {journal} {\bibinfo  {journal} {Phys. Fluids}\ }\textbf {\bibinfo
  {volume} {21}},\ \bibinfo {pages} {113301} (\bibinfo {year}
  {2009})}\BibitemShut {NoStop}%
\bibitem [{\citenamefont {Obligado}\ \emph {et~al.}(2014)\citenamefont
  {Obligado}, \citenamefont {Teitelbaum}, \citenamefont {Cartellier},
  \citenamefont {Mininni},\ and\ \citenamefont {Bourgoin}}]{obligado14}%
  \BibitemOpen
  \bibfield  {author} {\bibinfo {author} {\bibfnamefont {M.}~\bibnamefont
  {Obligado}}, \bibinfo {author} {\bibfnamefont {T.}~\bibnamefont
  {Teitelbaum}}, \bibinfo {author} {\bibfnamefont {A.}~\bibnamefont
  {Cartellier}}, \bibinfo {author} {\bibfnamefont {P.}~\bibnamefont {Mininni}},
  \ and\ \bibinfo {author} {\bibfnamefont {M.}~\bibnamefont {Bourgoin}},\
  }\bibfield  {title} {\enquote {\bibinfo {title} {Preferential concentration
  of heavy particles in turbulence},}\ }\href@noop {} {\bibfield  {journal}
  {\bibinfo  {journal} {J. Turbul.}\ }\textbf {\bibinfo {volume} {15}},\
  \bibinfo {pages} {293--310} (\bibinfo {year} {2014})}\BibitemShut {NoStop}%
\bibitem [{\citenamefont {Uhlmann}\ and\ \citenamefont
  {Chouippe}(2017)}]{uhlmann_clustering_2017}%
  \BibitemOpen
  \bibfield  {author} {\bibinfo {author} {\bibfnamefont {M.}~\bibnamefont
  {Uhlmann}}\ and\ \bibinfo {author} {\bibfnamefont {A.}~\bibnamefont
  {Chouippe}},\ }\bibfield  {title} {\enquote {\bibinfo {title} {Clustering and
  preferential concentration of finite-size particles in forced
  homogeneous-isotropic turbulence},}\ }\href@noop {} {\bibfield  {journal}
  {\bibinfo  {journal} {J. Fluid. Mech.}\ }\textbf {\bibinfo {volume} {812}},\
  \bibinfo {pages} {991--1023} (\bibinfo {year} {2017})}\BibitemShut {NoStop}%
\bibitem [{\citenamefont {Balk}\ \emph {et~al.}(2004)\citenamefont {Balk},
  \citenamefont {Falkovich},\ and\ \citenamefont {Stepanov}}]{balk2004growth}%
  \BibitemOpen
  \bibfield  {author} {\bibinfo {author} {\bibfnamefont {A.~M.}\ \bibnamefont
  {Balk}}, \bibinfo {author} {\bibfnamefont {G.}~\bibnamefont {Falkovich}}, \
  and\ \bibinfo {author} {\bibfnamefont {M.~G.}\ \bibnamefont {Stepanov}},\
  }\bibfield  {title} {\enquote {\bibinfo {title} {Growth of density
  inhomogeneities in a flow of wave turbulence},}\ }\href@noop {} {\bibfield
  {journal} {\bibinfo  {journal} {Phys. Rev. Lett.}\ }\textbf {\bibinfo
  {volume} {92}},\ \bibinfo {pages} {244504} (\bibinfo {year}
  {2004})}\BibitemShut {NoStop}%
\bibitem [{\citenamefont {Aartrijk}\ \emph {et~al.}(2008)\citenamefont
  {Aartrijk}, \citenamefont {Clercx},\ and\ \citenamefont
  {Winters}}]{aartrijk08}%
  \BibitemOpen
  \bibfield  {author} {\bibinfo {author} {\bibfnamefont {M.~van}\ \bibnamefont
  {Aartrijk}}, \bibinfo {author} {\bibfnamefont {H.~J.~H.}\ \bibnamefont
  {Clercx}}, \ and\ \bibinfo {author} {\bibfnamefont {K.~B.}\ \bibnamefont
  {Winters}},\ }\bibfield  {title} {\enquote {\bibinfo {title}
  {Single-particle, particle-pair, and multiparticle dispersion of fluid
  particles in forced stably stratified turbulence},}\ }\href@noop {}
  {\bibfield  {journal} {\bibinfo  {journal} {Phys. Fluids}\ }\textbf {\bibinfo
  {volume} {20}},\ \bibinfo {pages} {025104} (\bibinfo {year}
  {2008})}\BibitemShut {NoStop}%
\bibitem [{\citenamefont {Sujovolsky}\ \emph {et~al.}(2018)\citenamefont
  {Sujovolsky}, \citenamefont {Mininni},\ and\ \citenamefont
  {Rast}}]{sujovolsky_single-particle_2017}%
  \BibitemOpen
  \bibfield  {author} {\bibinfo {author} {\bibfnamefont {N.~E.}\ \bibnamefont
  {Sujovolsky}}, \bibinfo {author} {\bibfnamefont {P.~D.}\ \bibnamefont
  {Mininni}}, \ and\ \bibinfo {author} {\bibfnamefont {M.~P.}\ \bibnamefont
  {Rast}},\ }\bibfield  {title} {\enquote {\bibinfo {title} {Single-particle
  dispersion in stably stratified turbulence},}\ }\href@noop {} {\bibfield
  {journal} {\bibinfo  {journal} {Phys. Rev. Fluids}\ }\textbf {\bibinfo
  {volume} {3}},\ \bibinfo {pages} {034603} (\bibinfo {year}
  {2018})}\BibitemShut {NoStop}%
\bibitem [{\citenamefont {Sujovolsky}\ and\ \citenamefont
  {Mininni}(2019)}]{sujovolsky_vertical_2018}%
  \BibitemOpen
  \bibfield  {author} {\bibinfo {author} {\bibfnamefont {N.~E.}\ \bibnamefont
  {Sujovolsky}}\ and\ \bibinfo {author} {\bibfnamefont {P.~D.}\ \bibnamefont
  {Mininni}},\ }\bibfield  {title} {\enquote {\bibinfo {title} {Vertical
  dispersion of {Lagrangian} tracers in fully developed stably stratified
  turbulence},}\ }\href@noop {} {\bibfield  {journal} {\bibinfo  {journal}
  {Phys. Rev. Fluids}\ }\textbf {\bibinfo {volume} {4}},\ \bibinfo {pages}
  {014503} (\bibinfo {year} {2019})}\BibitemShut {NoStop}%
\bibitem [{\citenamefont {van Aartrijk}\ and\ \citenamefont
  {Clercx}(2010)}]{van_aartrijk_vertical_2010}%
  \BibitemOpen
  \bibfield  {author} {\bibinfo {author} {\bibfnamefont {M.}~\bibnamefont {van
  Aartrijk}}\ and\ \bibinfo {author} {\bibfnamefont {H.~J.~H.}\ \bibnamefont
  {Clercx}},\ }\bibfield  {title} {\enquote {\bibinfo {title} {Vertical
  dispersion of light inertial particles in stably stratified turbulence: {The}
  influence of the {Basset} force},}\ }\href@noop {} {\bibfield  {journal}
  {\bibinfo  {journal} {Phys. Fluids}\ }\textbf {\bibinfo {volume} {22}},\
  \bibinfo {pages} {013301} (\bibinfo {year} {2010})}\BibitemShut {NoStop}%
\bibitem [{\citenamefont {Sozza}\ \emph {et~al.}(2016)\citenamefont {Sozza},
  \citenamefont {De~Lillo}, \citenamefont {Musacchio},\ and\ \citenamefont
  {Boffetta}}]{sozza16}%
  \BibitemOpen
  \bibfield  {author} {\bibinfo {author} {\bibfnamefont {A.}~\bibnamefont
  {Sozza}}, \bibinfo {author} {\bibfnamefont {F.}~\bibnamefont {De~Lillo}},
  \bibinfo {author} {\bibfnamefont {S.}~\bibnamefont {Musacchio}}, \ and\
  \bibinfo {author} {\bibfnamefont {G.}~\bibnamefont {Boffetta}},\ }\bibfield
  {title} {\enquote {\bibinfo {title} {Large-scale confinement and small-scale
  clustering of floating particles in stratified turbulence},}\ }\href@noop {}
  {\bibfield  {journal} {\bibinfo  {journal} {Phys. Rev. Fluids}\ }\textbf
  {\bibinfo {volume} {1}},\ \bibinfo {pages} {052401} (\bibinfo {year}
  {2016})}\BibitemShut {NoStop}%
\bibitem [{\citenamefont {Sozza}\ \emph {et~al.}(2018)\citenamefont {Sozza},
  \citenamefont {De~Lillo},\ and\ \citenamefont
  {Boffetta}}]{sozza2018inertial}%
  \BibitemOpen
  \bibfield  {author} {\bibinfo {author} {\bibfnamefont {A.}~\bibnamefont
  {Sozza}}, \bibinfo {author} {\bibfnamefont {F.}~\bibnamefont {De~Lillo}}, \
  and\ \bibinfo {author} {\bibfnamefont {G.}~\bibnamefont {Boffetta}},\
  }\bibfield  {title} {\enquote {\bibinfo {title} {Inertial floaters in
  stratified turbulence},}\ }\href@noop {} {\bibfield  {journal} {\bibinfo
  {journal} {EPL (Europhys. Letters)}\ }\textbf {\bibinfo {volume} {121}},\
  \bibinfo {pages} {14002} (\bibinfo {year} {2018})}\BibitemShut {NoStop}%
\bibitem [{\citenamefont {Francois}\ \emph {et~al.}(2014)\citenamefont
  {Francois}, \citenamefont {Xia}, \citenamefont {Punzmann}, \citenamefont
  {Ramsden},\ and\ \citenamefont {Shats}}]{francois2014three}%
  \BibitemOpen
  \bibfield  {author} {\bibinfo {author} {\bibfnamefont {N.}~\bibnamefont
  {Francois}}, \bibinfo {author} {\bibfnamefont {H.}~\bibnamefont {Xia}},
  \bibinfo {author} {\bibfnamefont {H.}~\bibnamefont {Punzmann}}, \bibinfo
  {author} {\bibfnamefont {S.}~\bibnamefont {Ramsden}}, \ and\ \bibinfo
  {author} {\bibfnamefont {M.}~\bibnamefont {Shats}},\ }\bibfield  {title}
  {\enquote {\bibinfo {title} {Three-dimensional fluid motion in {F}araday
  waves: {C}reation of vorticity and generation of two-dimensional
  turbulence},}\ }\href@noop {} {\bibfield  {journal} {\bibinfo  {journal}
  {Phys. Rev. X}\ }\textbf {\bibinfo {volume} {4}},\ \bibinfo {pages} {021021}
  (\bibinfo {year} {2014})}\BibitemShut {NoStop}%
\bibitem [{\citenamefont {Punzmann}\ \emph {et~al.}(2014)\citenamefont
  {Punzmann}, \citenamefont {Francois}, \citenamefont {Xia}, \citenamefont
  {Falkovich},\ and\ \citenamefont {Shats}}]{punzmann2014generation}%
  \BibitemOpen
  \bibfield  {author} {\bibinfo {author} {\bibfnamefont {H.}~\bibnamefont
  {Punzmann}}, \bibinfo {author} {\bibfnamefont {N.}~\bibnamefont {Francois}},
  \bibinfo {author} {\bibfnamefont {H.}~\bibnamefont {Xia}}, \bibinfo {author}
  {\bibfnamefont {G.}~\bibnamefont {Falkovich}}, \ and\ \bibinfo {author}
  {\bibfnamefont {M.}~\bibnamefont {Shats}},\ }\bibfield  {title} {\enquote
  {\bibinfo {title} {Generation and reversal of surface flows by propagating
  waves},}\ }\href@noop {} {\bibfield  {journal} {\bibinfo  {journal} {Nature
  Phys.}\ }\textbf {\bibinfo {volume} {10}},\ \bibinfo {pages} {658} (\bibinfo
  {year} {2014})}\BibitemShut {NoStop}%
\bibitem [{\citenamefont {Denissenko}\ \emph {et~al.}(2006)\citenamefont
  {Denissenko}, \citenamefont {Falkovich},\ and\ \citenamefont
  {Lukaschuk}}]{denissenko2006waves}%
  \BibitemOpen
  \bibfield  {author} {\bibinfo {author} {\bibfnamefont {P.}~\bibnamefont
  {Denissenko}}, \bibinfo {author} {\bibfnamefont {G.}~\bibnamefont
  {Falkovich}}, \ and\ \bibinfo {author} {\bibfnamefont {S.}~\bibnamefont
  {Lukaschuk}},\ }\bibfield  {title} {\enquote {\bibinfo {title} {How waves
  affect the distribution of particles that float on a liquid surface},}\
  }\href@noop {} {\bibfield  {journal} {\bibinfo  {journal} {Phys. Rev. Lett.}\
  }\textbf {\bibinfo {volume} {97}},\ \bibinfo {pages} {244501} (\bibinfo
  {year} {2006})}\BibitemShut {NoStop}%
\bibitem [{\citenamefont {Thomson}\ \emph {et~al.}(2009)\citenamefont
  {Thomson}, \citenamefont {Gemmrich},\ and\ \citenamefont
  {Jessup}}]{thomson_energy_2009}%
  \BibitemOpen
  \bibfield  {author} {\bibinfo {author} {\bibfnamefont {J.}~\bibnamefont
  {Thomson}}, \bibinfo {author} {\bibfnamefont {J.~R.}\ \bibnamefont
  {Gemmrich}}, \ and\ \bibinfo {author} {\bibfnamefont {A.~T.}\ \bibnamefont
  {Jessup}},\ }\bibfield  {title} {\enquote {\bibinfo {title} {Energy
  dissipation and the spectral distribution of whitecaps},}\ }\href@noop {}
  {\bibfield  {journal} {\bibinfo  {journal} {Geophys. Res. Lett.}\ }\textbf
  {\bibinfo {volume} {36}} (\bibinfo {year} {2009})}\BibitemShut {NoStop}%
\bibitem [{\citenamefont {Garrett}\ and\ \citenamefont
  {Munk}(1975)}]{garrett_space-time_1975}%
  \BibitemOpen
  \bibfield  {author} {\bibinfo {author} {\bibfnamefont {C.}~\bibnamefont
  {Garrett}}\ and\ \bibinfo {author} {\bibfnamefont {W.}~\bibnamefont {Munk}},\
  }\bibfield  {title} {\enquote {\bibinfo {title} {Space-time scales of
  internal waves: {A} progress report},}\ }\href@noop {} {\bibfield  {journal}
  {\bibinfo  {journal} {J. Geophys. Res.}\ }\textbf {\bibinfo {volume} {80}},\
  \bibinfo {pages} {291--297} (\bibinfo {year} {1975})}\BibitemShut {NoStop}%
\bibitem [{\citenamefont {D'Asaro}\ and\ \citenamefont
  {Lien}(2000)}]{dasaro_lagrangian_2000}%
  \BibitemOpen
  \bibfield  {author} {\bibinfo {author} {\bibfnamefont {E.~A.}\ \bibnamefont
  {D'Asaro}}\ and\ \bibinfo {author} {\bibfnamefont {R.-C.}\ \bibnamefont
  {Lien}},\ }\bibfield  {title} {\enquote {\bibinfo {title} {Lagrangian
  measurements of waves and turbulence in stratified flows},}\ }\href@noop {}
  {\bibfield  {journal} {\bibinfo  {journal} {J. Phys. Oceanogr.}\ }\textbf
  {\bibinfo {volume} {30}},\ \bibinfo {pages} {641--655} (\bibinfo {year}
  {2000})}\BibitemShut {NoStop}%
\bibitem [{\citenamefont {Jacobs}\ \emph {et~al.}(2015)\citenamefont {Jacobs},
  \citenamefont {Huntley}, \citenamefont {Kirwan}, \citenamefont {Lipphardt},
  \citenamefont {Campbell}, \citenamefont {Smith}, \citenamefont {Edwards},\
  and\ \citenamefont {Bartels}}]{jacobs_ocean_2015}%
  \BibitemOpen
  \bibfield  {author} {\bibinfo {author} {\bibfnamefont {G.~A.}\ \bibnamefont
  {Jacobs}}, \bibinfo {author} {\bibfnamefont {H.~S.}\ \bibnamefont {Huntley}},
  \bibinfo {author} {\bibfnamefont {A.~D.}\ \bibnamefont {Kirwan}}, \bibinfo
  {author} {\bibfnamefont {B.~L.}\ \bibnamefont {Lipphardt}}, \bibinfo {author}
  {\bibfnamefont {T.}~\bibnamefont {Campbell}}, \bibinfo {author}
  {\bibfnamefont {T.}~\bibnamefont {Smith}}, \bibinfo {author} {\bibfnamefont
  {K.}~\bibnamefont {Edwards}}, \ and\ \bibinfo {author} {\bibfnamefont
  {B.}~\bibnamefont {Bartels}},\ }\bibfield  {title} {\enquote {\bibinfo
  {title} {Ocean processes underlying surface clustering},}\ }\href@noop {}
  {\bibfield  {journal} {\bibinfo  {journal} {J. Geophys. Res.: Oceans}\
  }\textbf {\bibinfo {volume} {121}},\ \bibinfo {pages} {180--197} (\bibinfo
  {year} {2015})}\BibitemShut {NoStop}%
\bibitem [{\citenamefont {Guti\'errez}\ and\ \citenamefont
  {Auma\^itre}(2016)}]{gutierrez_clustering_2016}%
  \BibitemOpen
  \bibfield  {author} {\bibinfo {author} {\bibfnamefont {P.}~\bibnamefont
  {Guti\'errez}}\ and\ \bibinfo {author} {\bibfnamefont {S.}~\bibnamefont
  {Auma\^itre}},\ }\bibfield  {title} {\enquote {\bibinfo {title} {Clustering
  of floaters on the free surface of a turbulent flow: {An} experimental
  study},}\ }\href@noop {} {\bibfield  {journal} {\bibinfo  {journal} {Eur. J.
  Mech. B Fluids}\ }\textbf {\bibinfo {volume} {60}},\ \bibinfo {pages}
  {24--32} (\bibinfo {year} {2016})}\BibitemShut {NoStop}%
\bibitem [{\citenamefont {D'Asaro}(2014)}]{dasaro_turbulence_2014}%
  \BibitemOpen
  \bibfield  {author} {\bibinfo {author} {\bibfnamefont {E.~A.}\ \bibnamefont
  {D'Asaro}},\ }\bibfield  {title} {\enquote {\bibinfo {title} {Turbulence in
  the upper-ocean mixed layer},}\ }\href@noop {} {\bibfield  {journal}
  {\bibinfo  {journal} {Annu. Rev. Mar. Sci.}\ }\textbf {\bibinfo {volume}
  {6}},\ \bibinfo {pages} {101--115} (\bibinfo {year} {2014})}\BibitemShut
  {NoStop}%
\bibitem [{\citenamefont {Pearson}\ and\ \citenamefont
  {Fox-Kemper}(2018)}]{pearson_log-normal_2018}%
  \BibitemOpen
  \bibfield  {author} {\bibinfo {author} {\bibfnamefont {B.}~\bibnamefont
  {Pearson}}\ and\ \bibinfo {author} {\bibfnamefont {B.}~\bibnamefont
  {Fox-Kemper}},\ }\bibfield  {title} {\enquote {\bibinfo {title} {Log-normal
  turbulence dissipation in global ocean models},}\ }\href@noop {} {\bibfield
  {journal} {\bibinfo  {journal} {Phys. Rev. Lett.}\ }\textbf {\bibinfo
  {volume} {120}},\ \bibinfo {pages} {094501} (\bibinfo {year}
  {2018})}\BibitemShut {NoStop}%
\bibitem [{\citenamefont {Howell}\ \emph {et~al.}(2000)\citenamefont {Howell},
  \citenamefont {Buhrow}, \citenamefont {Heath}, \citenamefont {McKenna},
  \citenamefont {Hwang},\ and\ \citenamefont
  {Schatz}}]{howell2000measurements}%
  \BibitemOpen
  \bibfield  {author} {\bibinfo {author} {\bibfnamefont {D.~R.}\ \bibnamefont
  {Howell}}, \bibinfo {author} {\bibfnamefont {B.}~\bibnamefont {Buhrow}},
  \bibinfo {author} {\bibfnamefont {T.}~\bibnamefont {Heath}}, \bibinfo
  {author} {\bibfnamefont {C.}~\bibnamefont {McKenna}}, \bibinfo {author}
  {\bibfnamefont {W.}~\bibnamefont {Hwang}}, \ and\ \bibinfo {author}
  {\bibfnamefont {M.~F.}\ \bibnamefont {Schatz}},\ }\bibfield  {title}
  {\enquote {\bibinfo {title} {Measurements of surface-wave damping in a
  container},}\ }\href@noop {} {\bibfield  {journal} {\bibinfo  {journal}
  {Phys. Fluids}\ }\textbf {\bibinfo {volume} {12}},\ \bibinfo {pages}
  {322--326} (\bibinfo {year} {2000})}\BibitemShut {NoStop}%
\bibitem [{\citenamefont {Campagne}\ \emph {et~al.}(2018)\citenamefont
  {Campagne}, \citenamefont {Hassaini}, \citenamefont {Redor}, \citenamefont
  {Sommeria}, \citenamefont {Valran}, \citenamefont {Viboud},\ and\
  \citenamefont {Mordant}}]{campagne2018impact}%
  \BibitemOpen
  \bibfield  {author} {\bibinfo {author} {\bibfnamefont {A.}~\bibnamefont
  {Campagne}}, \bibinfo {author} {\bibfnamefont {R.}~\bibnamefont {Hassaini}},
  \bibinfo {author} {\bibfnamefont {I.}~\bibnamefont {Redor}}, \bibinfo
  {author} {\bibfnamefont {J.}~\bibnamefont {Sommeria}}, \bibinfo {author}
  {\bibfnamefont {T.}~\bibnamefont {Valran}}, \bibinfo {author} {\bibfnamefont
  {S.}~\bibnamefont {Viboud}}, \ and\ \bibinfo {author} {\bibfnamefont
  {N.}~\bibnamefont {Mordant}},\ }\bibfield  {title} {\enquote {\bibinfo
  {title} {Impact of dissipation on the energy spectrum of experimental
  turbulence of gravity surface waves},}\ }\href@noop {} {\bibfield  {journal}
  {\bibinfo  {journal} {Phys. Rev. Fluids}\ }\textbf {\bibinfo {volume} {3}},\
  \bibinfo {pages} {044801} (\bibinfo {year} {2018})}\BibitemShut {NoStop}%
\bibitem [{\citenamefont {Zakharov}\ and\ \citenamefont
  {Filonenko}(1966)}]{zakharov66}%
  \BibitemOpen
  \bibfield  {author} {\bibinfo {author} {\bibfnamefont {V.~E.}\ \bibnamefont
  {Zakharov}}\ and\ \bibinfo {author} {\bibfnamefont {N.~N.}\ \bibnamefont
  {Filonenko}},\ }\bibfield  {title} {\enquote {\bibinfo {title} {Energy
  spectrum for stochastic oscillations of the surface of a liquid},}\
  }\href@noop {} {\bibfield  {journal} {\bibinfo  {journal} {Dokl. Akad. Nauk
  SSSR}\ }\textbf {\bibinfo {volume} {170}},\ \bibinfo {pages} {1292--1295}
  (\bibinfo {year} {1966})}\BibitemShut {NoStop}%
\bibitem [{\citenamefont {Bruno}\ and\ \citenamefont
  {Carbone}(2013)}]{bruno2013solar}%
  \BibitemOpen
  \bibfield  {author} {\bibinfo {author} {\bibfnamefont {Roberto}\ \bibnamefont
  {Bruno}}\ and\ \bibinfo {author} {\bibfnamefont {Vincenzo}\ \bibnamefont
  {Carbone}},\ }\bibfield  {title} {\enquote {\bibinfo {title} {The solar wind
  as a turbulence laboratory},}\ }\href@noop {} {\bibfield  {journal} {\bibinfo
   {journal} {Living Reviews in Solar Physics}\ }\textbf {\bibinfo {volume}
  {10}},\ \bibinfo {pages} {2} (\bibinfo {year} {2013})}\BibitemShut {NoStop}%
\bibitem [{\citenamefont {Hurst}\ and\ \citenamefont
  {Vassilicos}(2007)}]{hurst2007scalings}%
  \BibitemOpen
  \bibfield  {author} {\bibinfo {author} {\bibfnamefont {D}~\bibnamefont
  {Hurst}}\ and\ \bibinfo {author} {\bibfnamefont {JC}~\bibnamefont
  {Vassilicos}},\ }\bibfield  {title} {\enquote {\bibinfo {title} {Scalings and
  decay of fractal-generated turbulence},}\ }\href@noop {} {\bibfield
  {journal} {\bibinfo  {journal} {Physics of Fluids}\ }\textbf {\bibinfo
  {volume} {19}},\ \bibinfo {pages} {035103} (\bibinfo {year}
  {2007})}\BibitemShut {NoStop}%
\bibitem [{\citenamefont {Cortet}\ \emph {et~al.}(2009)\citenamefont {Cortet},
  \citenamefont {Diribarne}, \citenamefont {Monchaux}, \citenamefont
  {Chiffaudel}, \citenamefont {Daviaud},\ and\ \citenamefont
  {Dubrulle}}]{cortet2009normalized}%
  \BibitemOpen
  \bibfield  {author} {\bibinfo {author} {\bibfnamefont {Pierre-Philippe}\
  \bibnamefont {Cortet}}, \bibinfo {author} {\bibfnamefont {Pantxo}\
  \bibnamefont {Diribarne}}, \bibinfo {author} {\bibfnamefont {Romain}\
  \bibnamefont {Monchaux}}, \bibinfo {author} {\bibfnamefont {Arnaud}\
  \bibnamefont {Chiffaudel}}, \bibinfo {author} {\bibfnamefont
  {Fran{\c{c}}ois}\ \bibnamefont {Daviaud}}, \ and\ \bibinfo {author}
  {\bibfnamefont {B{\'e}reng{\`e}re}\ \bibnamefont {Dubrulle}},\ }\bibfield
  {title} {\enquote {\bibinfo {title} {Normalized kinetic energy as a
  hydrodynamical global quantity for inhomogeneous anisotropic turbulence},}\
  }\href@noop {} {\bibfield  {journal} {\bibinfo  {journal} {Physics of
  Fluids}\ }\textbf {\bibinfo {volume} {21}},\ \bibinfo {pages} {025104}
  (\bibinfo {year} {2009})}\BibitemShut {NoStop}%
\bibitem [{\citenamefont {Nicolleau}\ and\ \citenamefont
  {Vassilicos}(2000)}]{nicolleau_turbulent_2000}%
  \BibitemOpen
  \bibfield  {author} {\bibinfo {author} {\bibfnamefont {F.}~\bibnamefont
  {Nicolleau}}\ and\ \bibinfo {author} {\bibfnamefont {J.~C.}\ \bibnamefont
  {Vassilicos}},\ }\bibfield  {title} {\enquote {\bibinfo {title} {Turbulent
  diffusion in stably stratified non-decaying turbulence},}\ }\href@noop {}
  {\bibfield  {journal} {\bibinfo  {journal} {J. Fluid. Mech.}\ }\textbf
  {\bibinfo {volume} {410}},\ \bibinfo {pages} {123--146} (\bibinfo {year}
  {2000})}\BibitemShut {NoStop}%
\bibitem [{\citenamefont {Craik}\ and\ \citenamefont
  {Leibovich}(1976)}]{craik76}%
  \BibitemOpen
  \bibfield  {author} {\bibinfo {author} {\bibfnamefont {A.~D.~D.}\
  \bibnamefont {Craik}}\ and\ \bibinfo {author} {\bibfnamefont
  {S.}~\bibnamefont {Leibovich}},\ }\bibfield  {title} {\enquote {\bibinfo
  {title} {A rational model for {L}angmuir circulations},}\ }\href@noop {}
  {\bibfield  {journal} {\bibinfo  {journal} {J. Fluid. Mech.}\ }\textbf
  {\bibinfo {volume} {73}},\ \bibinfo {pages} {401--426} (\bibinfo {year}
  {1976})}\BibitemShut {NoStop}%
\bibitem [{\citenamefont {Craik}(1982)}]{craik82}%
  \BibitemOpen
  \bibfield  {author} {\bibinfo {author} {\bibfnamefont {A.~D.~D.}\
  \bibnamefont {Craik}},\ }\bibfield  {title} {\enquote {\bibinfo {title} {The
  drift velocity of water waves},}\ }\href@noop {} {\bibfield  {journal}
  {\bibinfo  {journal} {J. Fluid. Mech.}\ }\textbf {\bibinfo {volume} {116}},\
  \bibinfo {pages} {187--205} (\bibinfo {year} {1982})}\BibitemShut {NoStop}%
\bibitem [{\citenamefont {Andrews}\ and\ \citenamefont
  {McIntyre}(1978)}]{andrews78}%
  \BibitemOpen
  \bibfield  {author} {\bibinfo {author} {\bibfnamefont {D.~G.}\ \bibnamefont
  {Andrews}}\ and\ \bibinfo {author} {\bibfnamefont {M.~E.}\ \bibnamefont
  {McIntyre}},\ }\bibfield  {title} {\enquote {\bibinfo {title} {An exact
  theory of nonlinear waves on a {L}agrangian-mean flow},}\ }\href@noop {}
  {\bibfield  {journal} {\bibinfo  {journal} {J. Fluid. Mech.}\ }\textbf
  {\bibinfo {volume} {89}},\ \bibinfo {pages} {609--646} (\bibinfo {year}
  {1978})}\BibitemShut {NoStop}%
\bibitem [{\citenamefont {Holm}(1999)}]{holm99}%
  \BibitemOpen
  \bibfield  {author} {\bibinfo {author} {\bibfnamefont {D.~D.}\ \bibnamefont
  {Holm}},\ }\bibfield  {title} {\enquote {\bibinfo {title} {Fluctuation
  effects on 3{D} {L}agrangian mean and {E}ulerian mean fluid motion},}\
  }\href@noop {} {\bibfield  {journal} {\bibinfo  {journal} {Physica D}\
  }\textbf {\bibinfo {volume} {133}},\ \bibinfo {pages} {215--269} (\bibinfo
  {year} {1999})}\BibitemShut {NoStop}%
\bibitem [{\citenamefont {Aste}\ and\ \citenamefont
  {Di~Matteo}(2008)}]{aste08}%
  \BibitemOpen
  \bibfield  {author} {\bibinfo {author} {\bibfnamefont {T.}~\bibnamefont
  {Aste}}\ and\ \bibinfo {author} {\bibfnamefont {T.}~\bibnamefont
  {Di~Matteo}},\ }\bibfield  {title} {\enquote {\bibinfo {title} {Emergence of
  {G}amma distributions in granular materials and packing models},}\
  }\href@noop {} {\bibfield  {journal} {\bibinfo  {journal} {Phys. Rev. E}\
  }\textbf {\bibinfo {volume} {77}},\ \bibinfo {pages} {021309} (\bibinfo
  {year} {2008})}\BibitemShut {NoStop}%
\bibitem [{\citenamefont {Boffetta}\ \emph {et~al.}(2004)\citenamefont
  {Boffetta}, \citenamefont {Davoudi}, \citenamefont {Eckhardt},\ and\
  \citenamefont {Schumacher}}]{boffetta04}%
  \BibitemOpen
  \bibfield  {author} {\bibinfo {author} {\bibfnamefont {G.}~\bibnamefont
  {Boffetta}}, \bibinfo {author} {\bibfnamefont {J.}~\bibnamefont {Davoudi}},
  \bibinfo {author} {\bibfnamefont {B.}~\bibnamefont {Eckhardt}}, \ and\
  \bibinfo {author} {\bibfnamefont {J.}~\bibnamefont {Schumacher}},\ }\bibfield
   {title} {\enquote {\bibinfo {title} {Lagrangian tracers on a surface flow:
  {T}he role of time correlations},}\ }\href@noop {} {\bibfield  {journal}
  {\bibinfo  {journal} {Phys. Rev. Lett.}\ }\textbf {\bibinfo {volume} {93}},\
  \bibinfo {pages} {134501} (\bibinfo {year} {2004})}\BibitemShut {NoStop}%
\bibitem [{\citenamefont {Dalbe}\ \emph {et~al.}(2011)\citenamefont {Dalbe},
  \citenamefont {Cosic}, \citenamefont {Berhanu},\ and\ \citenamefont
  {Kudrolli}}]{dalbe11}%
  \BibitemOpen
  \bibfield  {author} {\bibinfo {author} {\bibfnamefont {M.-J.}\ \bibnamefont
  {Dalbe}}, \bibinfo {author} {\bibfnamefont {D.}~\bibnamefont {Cosic}},
  \bibinfo {author} {\bibfnamefont {M.}~\bibnamefont {Berhanu}}, \ and\
  \bibinfo {author} {\bibfnamefont {A.}~\bibnamefont {Kudrolli}},\ }\bibfield
  {title} {\enquote {\bibinfo {title} {Aggregation of frictional particles due
  to capillary attraction},}\ }\href@noop {} {\bibfield  {journal} {\bibinfo
  {journal} {Phys. Rev. E}\ }\textbf {\bibinfo {volume} {83}},\ \bibinfo
  {pages} {051403} (\bibinfo {year} {2011})}\BibitemShut {NoStop}%
\bibitem [{\citenamefont {Falkovich}\ \emph {et~al.}()\citenamefont
  {Falkovich}, \citenamefont {Weinberg}, \citenamefont {Denissenko},\ and\
  \citenamefont {Lukaschuk}}]{falkovich05}%
  \BibitemOpen
  \bibfield  {author} {\bibinfo {author} {\bibfnamefont {G.}~\bibnamefont
  {Falkovich}}, \bibinfo {author} {\bibfnamefont {A.}~\bibnamefont {Weinberg}},
  \bibinfo {author} {\bibfnamefont {P.}~\bibnamefont {Denissenko}}, \ and\
  \bibinfo {author} {\bibfnamefont {S.}~\bibnamefont {Lukaschuk}},\ }\bibfield
  {title} {\enquote {\bibinfo {title} {Floater clustering in a standing
  wave},}\ }\href@noop {} {\bibfield  {journal} {\bibinfo  {journal} {Nature}\
  }\textbf {\bibinfo {volume} {435}},\ \bibinfo {pages}
  {1045--1046}}\BibitemShut {NoStop}%
\end{thebibliography}%

\end{document}